\documentclass[aps,twocolumn,floatfix]{revtex4-1}
\usepackage{times}
\usepackage{graphics}
\usepackage{grffile} 
\usepackage{epsfig,calc,color,xspace}

\usepackage{bm}

\usepackage{placeins} 
\usepackage[breaklinks=true,
            pdfborder={0 0 1},
            colorlinks=true,
            linkcolor=black,
            citecolor=blue,
            urlcolor=blue]{hyperref}

\usepackage{amsmath,amssymb}
\usepackage{chemarr} 
\usepackage[normalem]{ulem}
\usepackage{verbatim}  

\newcommand{\mfh}[1]{\textcolor{blue}{#1}}

\newcommand{\ie}{\textit{i.e.}\xspace}
\newcommand{\eg}{\textit{e.g.}\xspace}
\newcommand{\etal}{et al.\xspace}

\newcommand{\NC}{`No Allostery'\xspace}
\newcommand{\NA}{`No Induced Fit'\xspace}
\newcommand{\YA}{`Induced Fit'\xspace}

\newcommand{\kt}{k_{\text{B}}T}
\newcommand{\pintt}{P_\text{int}}
\newcommand{\pint}{P_\text{int}}
\newcommand{\fc}{P_\text{N}}
\def\ptrap{p_\text{trap}}
\newcommand{\halftime}{\tau_\text{1/2}}
\newcommand{\nnuc}{n_\text{nuc}}
\newcommand{\nn}{{\hat{n}}}
\newcommand{\nb}{n_\text{B}}
\newcommand{\ckt}{c_\text{kt}}
\newcommand{\cc}{c_\text{c}}
\newcommand{\cmin}{c_\text{min}}

\newcommand{\telong}{\tau_\text{elong}}

\newcommand{\tnuc}{\tau_\text{nuc}^\text{min}}
\newcommand{\tmax}{\tau_\text{max}}
\newcommand{\fa}{f_\text{A}}
\newcommand{\gA}{g_\text{A}}
\newcommand{\eb}{\epsilon_\text{b}}
\newcommand{\ssb}{s_\text{b}}
\newcommand{\ssc}{s_\text{c}}
\newcommand{\kb}{k_\text{B}}
\newcommand{\gb}{g_\text{b}}
\newcommand{\dgn}{\Delta g_n}  
\newcommand{\gkt}{g_\text{b}^\text{kt}}  
\newcommand{\gbeff}{g_\text{b}^\text{shift}}  
\newcommand{\ebeff}{\epsilon_\text{b}^\text{eff}}  
\newcommand{\gnuc}{\gb}  

\newcommand{\gelong}{g_{\text{elong}}}
\newcommand{\Ahalf}{A_{1/2}}  
\newcommand{\cratio}{\delta c}

\newcommand{\nci}{n_{\text{c,}n}}
\newcommand{\gN}{g_\text{N}}
\newcommand{\cstar}{c^{*}}

\def\kbar{{\bar{k}}}

\def\Ka{K_\text{A}}
\def\mc{m_\text{c}}
\def\mA{m_\text{A}}
\def\gsub{g_\text{sub}}
\def\fActive{k_\text{A}}  
\def\bActive{\bar{k}_\text{A}}  

\newcommand{\rc}{r_\mathrm{c}}
\newcommand{\tc}{\theta_\mathrm{c}}
\newcommand{\pc}{\phi_\mathrm{c}}
\newcommand{\LJ}[1]{ \mathcal{L}_{#1} }

\def\btt1{{\tt$\backslash$\string1}}%

\def\AmS{{\protect\the\textfont2
        A\kern-.1667em\lower.5ex\hbox{M}\kern-.125emS}}

\citeindextrue

\begin{document}

\title{Allosteric control in icosahedral capsid assembly}
\author{Guillermo R. Lazaro}
\author{Michael F. Hagan}
\email{hagan@brandeis.edu}
\affiliation{Martin Fisher School of Physics, Brandeis University, Waltham, MA, 02454}
\date{\today}
\begin{abstract}
During the lifecycle of a virus, viral proteins and other components self-assemble to form an ordered protein shell called a capsid. This assembly process is subject to multiple competing constraints, including the need to form a thermostable shell while avoiding kinetic traps. It has been proposed that viral assembly satisfies these constraints through allosteric regulation, including the interconversion of capsid proteins among conformations with different propensities for assembly. In this article we use computational and theoretical modeling to explore how such allostery affects the assembly of icosahedral shells. We simulate assembly under a wide range of protein concentrations, protein binding affinities, and two different mechanisms of allosteric control. We find that, above a threshold strength of allosteric control, assembly becomes robust over a broad range of subunit binding affinities and concentrations, allowing the formation of highly thermostable capsids. Our results suggest that allostery can significantly shift the range of protein binding affinities that lead to successful assembly, and thus should be accounted for in models that are used to estimate interaction parameters from experimental data.
\end{abstract}

\maketitle


\section{Introduction}

The assembly of a virus outer protein shell (capsid) requires a delicate balance among thermodynamic and kinetic constraints. The proteins must assemble quickly to evade proteolysis and detection by the host, and their capsid must be sufficiently thermostable to survive intact under potentially harsh conditions while searching for a new infection target. Yet, self-assembly of ordered structures usually requires weak, reversible interactions among the components, since strong interactions lead to kinetic traps \cite{Whitelam2015,Hagan2014,Zlotnick2003}. Many viruses must also control the time and place of assembly, so that the capsid can select specific components from amidst a crowded cellular environment. A number of strategies have been proposed by which viruses control their assembly process to ensure productive and timely capsid formation (\eg \cite{Rao2006,Garmann2015,Perlmutter2015,Stockley2013b,Caspar1980}). One such strategy is allosteric regulation \cite{Packianathan2010,Caspar1980}, in which capsid proteins sample an ensemble of conformational states with different propensities for assembly, with the relative populations of different states influenced by binding of proteins or other molecules. In this article, we theoretically and computationally examine how allostery at the level of protein-protein interactions can lead to self-regulation of assembly kinetics.

Learning mechanistic information such as allosteric regulation from experiments alone is challenging because most assembly intermediates are transient, and thus not readily observed. For example, a wealth of information has been obtained from in vitro experiments in which capsid assembly kinetics are monitored by size exclusion chromatography (SEC) or x-ray or light scattering (e.g.) \cite{Prevelige1993,Zlotnick1999,Zlotnick2000,Casini2004,Chen2008,Berthet-Colominas1987,Kler2012,Kler2013}. However, since intermediates are usually undetectable, these bulk techniques primarily report on the concentrations of assembled capsids and unassembled subunits.
Recently, techniques which monitor individual capsids or can detect transient intermediates have begun to address this limitation \cite{Stockley2007,Basnak2010,Uetrecht2011,Baumgaertel2012,Jouvenet2011,Borodavka2012,Tresset2013,Pierson2014,Zhou2011,Law-Hine2015}. However, even these techniques provide structural data at limited resolution and cannot characterize the full ensemble of intermediates. Therefore, theoretical models are needed to obtain a complete understanding of capsid assembly from such experimental data.

Theoretical models have already played an important role in relating experimental data to assembly pathways and the driving forces that control them. For example, binding affinities have been estimated by fitting the ratio of assembled capsids to unassembled subunits measured at long times to the equilibrium law of mass action \cite{Ceres2002,Zlotnick2013}. Assembly kinetics have been analyzed using models in which capsid formation is viewed as the assembly of rigid subunits (lacking internal degrees of freedom) into polyhedral shells. These models can be formulated as a master equation and solved numerically (\eg \cite{Zlotnick1994,Zlotnick1999,Endres2002,Moisant2010,Schoot2007}) or analyzed by stochastically generating trajectories consistent with the master equation (\eg  \cite{Zhang2006,Keef2006,Hemberg2006,Dykeman2013a,Smith2014,Jamalyaria2005,Sweeney2008}).
Despite simplifications used to make these methods tractable, they capture many features of experimental assembly kinetics. Fitting their results against light scattering data has enabled estimates of physical parameters such as subunit-subunit binding affinities and rate constants \cite{Chen2008,Zlotnick1999,Zlotnick2011,Kumar2010,Xie2012}. These results suggest that capsid subunit-subunit binding affinities are generically weak, on the order of 4 kcal/mol \cite{Ceres2002,Zlotnick2003}. When binding affinities exceed this limit, assembly is limited by kinetic traps, in which the formation of long-lived disordered or partially assembled structures inhibit capsid formation.



Despite these important insights, current models do not capture all aspects of experimental data. Typically, models cannot quantitatively reproduce kinetics at all timescales across a range of concentrations \cite{Zlotnick2011}. Similarly, capsid disassembly exhibits a surprising degree of hysteresis considering the measured weakness of subunit binding affinities \cite{Singh2003,Roos2010}.
 For example, assembled capsids are highly stable in infinite dilution; \eg hepatitis B virus (HBV) capsids exhibit virtually no subunit dissociation even on a timescale of months \cite{Uetrecht2010a}. While this observation could be accounted for by a post-assembly maturation process that increases capsid stability, as yet there is no evidence for this in HBV.

One feature which has not yet been incorporated in most assembly models is that capsid proteins sample multiple conformational states, with different propensities for assembly. For example, structural studies on HBV \cite{Packianathan2010}, brome mosaic virus (BMV) and HIV \cite{Deshmukh2013} find that their capsid proteins adopt conformations in solution that are incompatible with insertion into a capsid, suggesting that the protein's primary conformation in solution is inactive for assembly (assembly-inactive), and that capsid formation requires a transition to an assembly-active conformation.  These observations have led to the hypothesis that assembly of icosahedral viruses may be subject to allosteric regulation \cite{Birnbaum1990,Ceres2002,Stray2005,Wingfield1995,Packianathan2010,Zlotnick2011,Deshmukh2013}.

It has long been appreciated that protein conformational switching can play a key role in the kinetics of assembling filamentous structures. Asakura \cite{Asakura1968} showed that the elongation rate of the bacterial flagellum is limited by the rate of flagellin monomer undergoing a conformational transition to its bound form. From this observation, he deduced that interaction with the flagellum triggers the conformational transition in the monomer by an `induced fit'. Similar observations were made for other filamentous assemblies, including the bacteriophage T4 tail \cite{Kikuchi1975} and tobacco mosaic virus \cite{Butler1999,Klug1999,Caspar1990,Kraft2012}. Caspar coined the term `autostery' \cite{Caspar1980,Caspar1991} to describe such an induced fit process, in which a protein that exists in an assembly-inactive conformation in solution is driven to switch to an assembly-active conformation by interacting with other copies of itself within an assemblage.  By controlling the rate of nucleation in comparison to elongation, autostery could provide  mechanism for self-regulation of assembly.

The role of an assembly-active/inactive transition in the assembly kinetics of icosahedral viruses has received less attention; more work has focused on the roles of conformational switching in overcoming the geometrical constraints imposed by an icosahedral geometry \cite{Caspar1962,Berger1994,Schwartz1998,Schwartz2000} and structural polymorphism \cite{Elrad2008,Nguyen2007,Nguyen2009,Basnak2010,Morton2010}. 
Although Packianathan \etal \cite{Packianathan2010} was primarily an experimental study, they also used a rate equation approach to compare the behaviors of assembly with no allosteric regulation, with induced fit (autostery), or allostery without induced fit.  They concluded that the induced fit mechanism leads to productive assembly and could increase  hysteresis associated with disassembly, but that allostery without induced fit did not lead to productive assembly, because such strong subunit affinities were required that kinetic trapping resulted.

In this article we perform a more extensive theoretical and computational investigation of the effects of allostery on assembly kinetics and their sensitivity to kinetic trapping.
In contrast to Ref.~\cite{Packianathan2010}, we find that both mechanisms of allosteric regulation (with or without induced fit) can drive productive assembly, although induced fit allows for productive assembly over a wider parameter range.
Under moderate parameter values, allostery does not enhance assembly robustness --- the width of the range of subunit concentrations or binding affinities leading to productive assembly is not increased. However, under sufficiently strong allosteric control (meaning that the population of unassembled subunits is strongly shifted toward the assembly-inactive conformation), high assembly yields are achieved over  a broad range of parameters, including high binding affinities. Our results highlight the importance of accounting for allostery in models used to estimate parameters from experimental data.

%
\begin{figure*}[hbt]
\begin{center}
  \includegraphics[width=.7\textwidth]{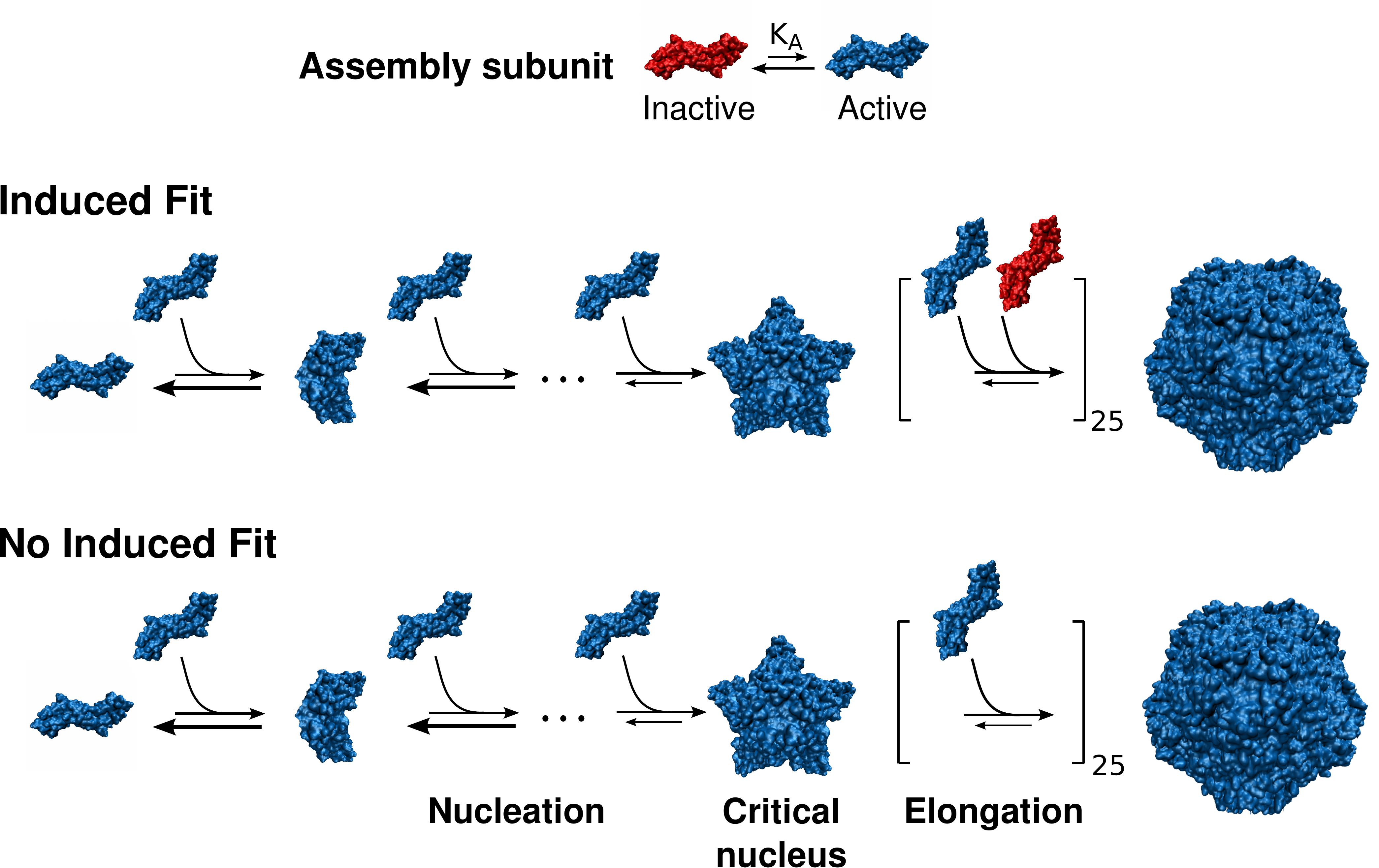}
     \caption{The two mechanisms of allostery that we consider in this article. The irreducible assembly unit (a protein dimer in this schematic) interconverts between assembly-inactive and assembly-active conformations, with equilibrium constant $\Ka$. In both mechanisms, only assembly-active subunits can combine to form intermediates smaller than a critical nucleus. In the \YA mechanism, both assembly-active and assembly-inactive subunits can bind to larger intermediates, whereas in the \NA mechanism, only assembly-active subunits can bind throughout the assembly process.}
\label{fig:scheme}
\end{center}
\end{figure*}

This paper is organized as follows. In the next section, we incorporate allostery into a computational model and a master equation description of assembly. In section `Scaling estimates for the effect of allostery on assembly timescales'
 we develop scaling estimates for the effects of allostery on assembly timescales and assembly robustness. Then, in the Results
 we test these scaling estimates against numerical results from the computational and master equation models. In the Discussion
  we summarize the key observations, identify potential further extensions to the models, and discuss implications for estimating parameters by fitting against experimental data. Finally, the appendix provides further details about the models and compares the computational and master equation models.

\section{Models}
\label{sec:models}

\subsection{Allostery Models}
\label{sec:allostery_models}

To represent allostery in our models, unassembled subunits interconvert between assembly-inactive and active conformations, with equilibrium constant $\Ka=\exp(-\gA/\kt)$ and $\gA>0$ the unfavorable free energy associated with a monomer adopting the active state. The case of no allostery corresponds to the limit $\Ka\rightarrow\infty$. For simplicity, here and throughout, we focus on the limit where the active/inactive interconversion rate is fast relative to assembly timescales. We discuss effects of a slow conformational interconversion step in the Discussion. 

The two allostery behaviors that we consider throughout the paper are illustrated by the schematics in Fig.~\ref{fig:scheme}.  In the first, there is allostery but no induced fit, meaning that only subunits in the active conformation can associate with any size intermediate. In the second, subunits in the inactive conformation can associate with intermediates equal to or larger than a nucleus comprising $\nnuc$ subunits.  For notational consistency, we refer to these cases respectively as \NA and \YA, and the case without allostery (all subunits are active) as \NC.

Structural evidence suggests that the active-inactive transition in the HBV protein is closely linked to formation of a stable critical nucleus \cite{Packianathan2010}. The latter is associated with the geometry of an icosahedral shell. Since smaller intermediates have fewer interactions per subunit than large ones, stability increases with intermediate size. The critical nucleus refers to the smallest intermediate (or ensemble of intermediates) from which assembly into a complete capsid is more probable than complete dissociation. In vitro experiments on several viruses have determined critical nucleus sizes corresponding to small polygons (\eg a pentamer or trimer of dimers) \cite{Hagan2014}. Similarly, in the computational model described below we find that under certain conditions the critical nucleus corresponds to the smallest polygon that can form, which is a pentagon as shown in Fig.~\ref{fig:patchy_model}C (see also Refs.~\cite{Hagan2010,Zandi2006}).
 Therefore, we consider the critical nucleus as the minimum seed capable of driving a conformation change. I.e., inactive subunits can bind to an intermediate at or above the critical nucleus size, but only active subunits can associate to pre-critical intermediates (Fig.~\ref{fig:scheme}).

We now describe the two modeling approaches we use to study the effects of allostery on assembly.

\subsection{Computational model}
\label{sec:simulations}
\begin{figure}[hbt]
  \begin{center}
  \includegraphics[width=\columnwidth]{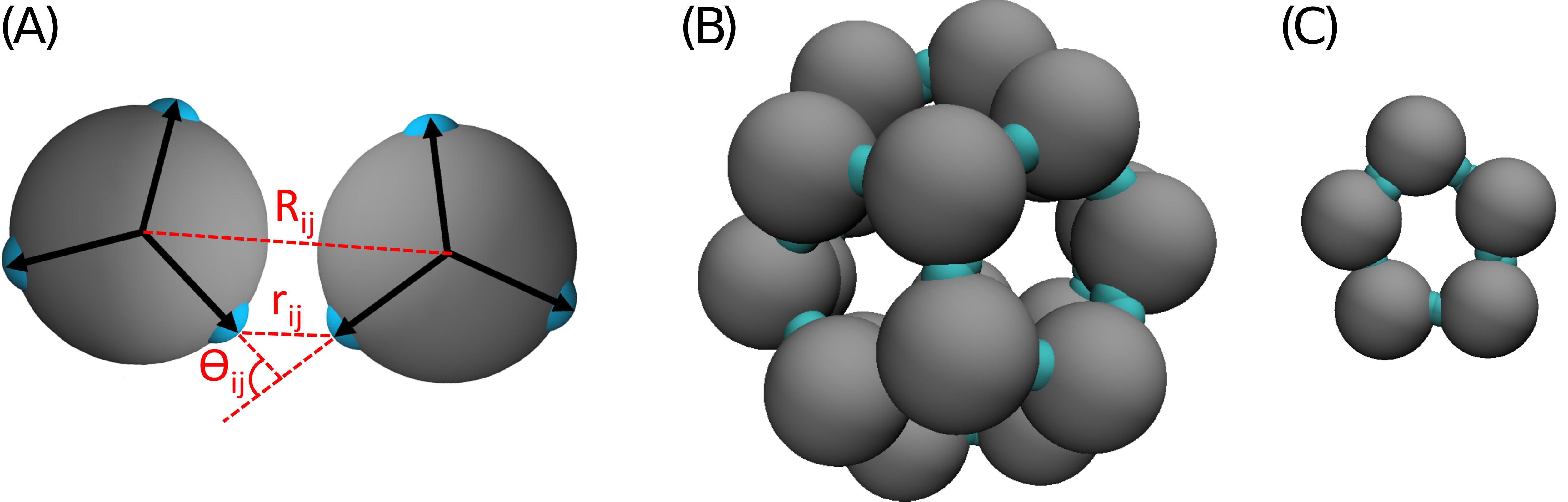}
    \caption{Computational model geometry. {\bf (A)} Geometry of two interacting subunits with the bond vectors depicted as arrows and attractors colored teal.  The angle between each of the subunit bond vectors is $108^{\mathrm{o}}$, and the interactions are described in the Appendix (Eqs. \eqref{eq_potential}).  The dihedral angle $\phi_{ij}^b$ is not shown. {\bf (B)} The complete capsid, which contains 20 subunits arranged with icosahedral symmetry. {\bf (C)} Critical nucleus for binding of assembly-inactive subunits in the \YA model.}
  \label{fig:patchy_model}
  \end{center}
\end{figure}

We consider a model for the assembly of icosahedral shells used in Ref.~\cite{Perkett2014}, in which subunits are modeled as rigid bodies,  with excluded volume interactions represented by spherically symmetric repulsive forces, and the complementary subunit-subunit interactions that drive assembly represented by directional attractions.  The lowest energy states in the model correspond to `capsids', which consist of $N=20$ subunits (each of which could represent a protein trimer) in a shell with icosahedral symmetry.  Because the spatial positions and orientations of all subunits are explicitly tracked, there are no assumptions about assembly pathways or the structures that emerge from assembly.

Following the approach developed by Schwartz et al. \cite{Schwartz1998}, the subunit excluded volume is spherically symmetric and three attractive patches (bond vectors) are rigidly fixed to the subunit, with each pair of bond vectors forming an angle of $108^\circ$ (see Fig. \ref{fig:patchy_model} and Eq.~\eqref{eq_potential}).  There is a favorable interaction between subunits when (1) the ends of bond vectors nearly overlap, (2) the bond vectors are nearly anti-parallel, and (3) the secondary bond vectors are nearly coplanar.  Twenty subunits realizing these conditions results in the minimum energy target structure (a complete capsid) shown in Fig. \ref{fig:patchy_model}. The interaction strength is tuned by the parameter $\eb$. All interaction potentials are pairwise.

The subunit-subunit interaction follows additional rules depending on the type of allostery being modeled. For \NC, all pairs of subunits meeting the interaction criteria listed above experience attractive interactions. For \NA, subunits stochastically switch between inactive and active conformations. Only pairs of active subunits satisfying the binding criteria experience attractive interactions. Since there is no autostery in this case, the effects of allostery on assembly arise only due to the cooperativity of capsid assembly.
For \YA, we define the threshold for autostery activity to be the formation of at least one polygon (any closed cycle of subunit-subunit interactions, such as a pentamer). Thus, in this case interactions are possible for any pair of active subunits, and also between inactive subunits and subunits within a partial capsid which has at least one complete polygon.
Full details of the model are given in the appendix.  

{\it Simulation parameters.} The parameters of the model are the energy associated with the attractive potential, $\eb$, and the specificity of the directional attractions, which is controlled by the angular parameters  $\tc$ and $\pc$. Subunit positions and orientations were propagated using overdamped Brownian dynamics according to a second order predictor-corrector algorithm~\cite{Branka1999,Heyes2000}, with the unit of time $t_0 = \sigma^2/D$, where $D$ is the subunit diffusion coefficient and $\sigma$ is the subunit diameter. We simulated systems with 500 subunits in a periodic box with side length 17, where all distances are measured in units of the subunit diameter $\sigma$. For each parameter set, results were averaged over 20 or more independent simulations. 
The orientational specifity parameters were  $\tc=0.5$ and $\pc=\pi$. These parameters tend to disfavor the formation of incorrect subunit-subunit interaction geometries and thus inhibit formation of malformed capsids, allowing us to study the high affinity limit. However, such configurations do arise at higher binding energies as discussed below. To obtain dimensionless units, we rescale energies by $\kt$ and times by  $t_0$. The subunit conformational switching rate was $2.5/t_0$.
Simulations were initialized by generating random positions and orientations for subunits, with subunit positions that led to subunit-subunit overlap (positive potential energies in excess of 1 $\kt$) rejected.

\subsection{Master equation model for capsid assembly}
\label{sec:rate_equations}
We also consider a master equation description of
polyhedral shell assembly, which is sufficiently computationally tractable to allow modeling assembly over broad parameter ranges. Specifically, we extend the `nucleation and growth' model described in Ref.\cite{Hagan2010} to include multiple subunit conformations. This model was based on the work of Zlotnick and coworkers \cite{Zlotnick1994,Zlotnick1999,Endres2002}, and consists of a system of coupled rate equations that describe the time evolution of concentrations of empty capsid intermediates:
\begin{eqnarray}
\frac{d c_1}{d t} &=& -k_1 c_1^2 + 2 \kbar_2 c_2 +\sum_{n=2}^{N}-k_n c_n c_1 + \kbar_n c_n \nonumber \\
\frac{d c_n}{d t}&=&k_{n-1} c_1 c_{n-1} - k_n c_1 c_n \qquad \qquad n=2\dots N \nonumber \\
& & -\kbar_n c_n + \kbar_{n+1} c_{n+1}
\label{eq:rateEquations}
\end{eqnarray}
where $c_n$ is the concentration of intermediates with $n$ subunits, and $k_n$ and $\kbar_n$ are respectively association and dissociation rate constants for intermediate $n$. The initial condition is $c_1(0)=c_0$,  $c_n(0)=0$ for $n>1$. The extensions of Eq.~\eqref{eq:rateEquations} to describe allostery  are given in the appendix. 

There are several important assumptions underlying the master equation: malformed capsids are not considered \cite{Schwartz1998,Hagan2006,Nguyen2007,Wilber2007,Hicks2006,Elrad2008,Rapaport2008,Nguyen2009}, assembly proceeds along a single pathway \cite{Endres2005,Zhang2006,Sweeney2008}, only single subunits can bind or unbind, and only one $k_n$ and $\kbar_n$ are considered for each size $n$ (averaged over all intermediates of that size). However, these assumptions are not present in the BD simulations described above,and we find close agreement between the two approaches (see appendix Fig.~\ref{fig:compare}). Moreover, Ref.~\cite{Hagan2010} showed that extending Eqs.~\eqref{eq:rateEquations} to relax these simplifications does not qualitatively change their predictions. Most importantly, rate equations of this form capture many features of experimental assembly kinetics data (\eg \cite{Zlotnick1999,Johnson2005}).

The association and dissociation rate constants are related by detailed balance $\kbar_n=k_n \exp(\dgn/\kt)/v_0$, with $\dgn=G_n-G_{n-1}$ the change in free energy due to association of a subunit. Association free energies $\dgn$, which can be fit to experimental data using the law of mass action \cite{Ceres2002,Ceres2004,Zlotnick2007}, include hydrophobic and electrostatic interactions \cite{Kegel2004} and depend on $p$H and salt concentration \cite{Ceres2002}. Specifying the assembly model requires defining the set of intermediates $n$ and the transition rates $\{k_n,\kbar_n\}$.

The model we use here is based on those of Zlotnick and coworkers \cite{Zlotnick1994,Zlotnick1999,Endres2002}, in which the subunit-subunit association free energy for intermediate $n$ is proportional to the number of new subunit-subunit contacts $\nci$ formed by addition of a subunit to that intermediate \cite{entropyNote}. Specifying $\{\dgn\}$ thus requires defining the geometry of each intermediate. This usually begins with specifying the geometry of a capsid and its subunits in terms of a polyhedron (for example, see Fig.~\ref{fig:patchy_model} or Fig.~1 in Ref.~\cite{Endres2002}), and assuming that assembly proceeds along a single path.  The path can be comprised of the lowest energy intermediate for each size $n$ \cite{Zlotnick1994} or correspond to an `average' pathway \cite{Endres2002} in which all subunits, except during the initial and final stages of assembly, make the same average number of contacts. We choose the latter definition, since it is simpler and both definitions lead to a qualitatively similar behavior. Specifically, the association rate constant $k$ is independent of intermediate size and association free energies are given by $\dgn=\gnuc$ before nucleation ($n<\nnuc$) and $\dgn=\gelong$ during elongation ($\nnuc \le n < N-1$), where $\nnuc$ is the critical nucleus size. Finally, inserting the last subunit makes the maximum number of possible contacts and thus enjoys the most favorable association free energy $\gN$.

Because this model was previously explored extensively in the absence of conformation changes \cite{Hagan2010}, we focus on one set of interaction parameters (except where noted otherwise):  capsid size $N=120$ corresponding to 120 dimer subunits in hepatitis B virus \cite{Ceres2002}, critical nucleus size $\nnuc=5$ (a pentamer of dimers), association rate constant $k=10^5$ M$^{-1}$s$^{-1}$ \cite{Johnson2005}, and free energy parameters $\gnuc=7\kt$ ($\approx 4$ kcal/mol) \cite{Ceres2002}, $\gelong=2\gnuc$ and $\gN=2\gelong$, which imply that adding a subunit becomes on average twice as favorable after nucleation and four times as favorable for the final subunit. 

\section{Scaling Estimates for the Effect Of Allostery on Assembly Timescales}
\label{sec:scaling}

To gain an intuitive understanding of how allostery can affect assembly, in this section we derive simple scaling estimates for the timescales associated with the two assembly mechanisms shown in Fig.~\ref{fig:scheme}, based on the master equation model (Eqs.~\ref{eq:rateEquations}). Although we introduce a number of simplifications, in the next section we show that the resulting scaling estimates apply at least qualitatively when these simplifications are relaxed in the computational and theoretical models. We closely follow Ref.~\cite{Hagan2010}, except that we extend the analysis to include allostery.

We consider a system of capsid protein subunits with total concentration $c_0$ that assemble into capsids with $N$ subunits. The word subunit refers to the basic assembly unit, which could be a protein dimer or larger oligomer \cite{Hagan2014}.
As in the master equation model, we break the assembly of a capsid into `nucleation' and `elongation' phases.  For simplicity we assume that the association rate constant $k$ is independent of intermediate size, so that for the \NC reaction rates of association to each intermediate are $k c_1$ with $c_1$ the concentration of free subunits.  We assume the limit of fast conformational interconversion, so
for the \NA case,  association rates are given by  $k \fa c_1$ with $\fa=\Ka/\left(1+\Ka\right)$.
For \YA, association rates are $k \fa c_1$ for intermediate size $n<\nnuc$ and $k c_1$ for $n\ge\nnuc$.

  We now write the time required for an individual capsid to assemble as $\tau = \tau_\text{nuc} + \telong$ with $\tau_\text{nuc}$ and $\telong$ the average times for nucleation and growth, respectively.
The elongation timescale can be estimated by the mean first passage time for a biased random walk with a reflecting boundary conditions at $\nnuc$ and absorbing boundary conditions at $N$, with forward and reverse hopping rates given by the subunit association and dissociation rates respectively \cite{Hagan2010}. For early in the reaction, when $c_1\approx c_0$, this results in
\begin{align}
\telong & \cong N / k c_0 \fa^{\mA}
\label{eq:telong}
\end{align}
where $\mA=1$ for \NA and $\mA=0$ for \YA. We see that the elongation time is equal for \NC and \YA since in the latter case all subunits can bind to post-nucleated intermediates. Ref.~\cite{Hagan2010} showed that the duration of the lag phase in light scattering is proportional to $\telong$, thus predicting that the lag time should scale inversely with subunit concentration. This prediction was recently confirmed by experiments on HBV assembly \cite{Selzer2014}.

The mean nucleation time at the beginning of the reaction can be estimated from the statistics of a random walk biased toward disassembly \cite{Hagan2010}. Including conformation dynamics results in $\tnuc \approx  k^{-1}\exp\left(G_{\nn} /\kt \right) c_0^{-\nnuc} \fa^{-\nnuc}$,
where $\nn=\nnuc-1$ so that $G_{\nn}$ is the interaction free energy of the structure just below the critical nucleus.
 This estimate can be understood by noting that the pre-critical nucleus is present with concentration $c_{\nn}\cong \exp(G_{\nn}) c_0^{\nn}\fa^{\nn}$, and the rate of active subunits associating to the precritical nucleus is given by $k c_0 \fa$.

However, because free subunits are depleted by assembly, the nucleation rate never reaches this value, and net nucleation asymptotically approaches zero as the concentration of completed capsids approaches its equilibrium value.  Thus, we estimate the median assembly time $\halftime$ (the time at which the reaction is 50\% complete) by treating the system as a two-state reaction with $\nnuc$-th order kinetics, which yields \cite{Hagan2010}
\begin{align}
\halftime \cong  \frac{\Ahalf \fc}{N k}  \exp\left(G_{\nn}/\kt\right) c_0^{-\nn} \fa^{-\nn-1}
\label{eq:halftime}
\end{align}
with $\Ahalf= \frac{2^{\nn}-1}{\nn}$, and $\fc$ as the equilibrium fraction of subunits in complete capsids.  The factor of $N^{-1}$ in Eq.~\ref{eq:halftime} accounts for the fact that $N$ subunits are depleted by each assembled capsid.

When capsid growth is fast compared to nucleation, the expressions  Eq.~\ref{eq:telong} and Eq.~\ref{eq:halftime} respectively predict the duration of the lag phase and the median assembly time.  However, these relations begin to fail as intermediate concentrations build up above a crossover concentration $\cc$ at which the initial elongation and nucleation times are equal
\begin{align}
 \cc \cong N^{-\frac{2}{\nn}} e^{\frac{ G_{\nn}}{\nn \kt}} \fa^{-\frac{\nn-\mA+1}{\nn}}
\label{eq:cc}
\end{align}
Significant kinetic trapping then sets in above a threshold concentration set by $\telong=\halftime$:
\begin{align}
 \ckt \cong \left(\Ahalf N^{-2}\right)^{\frac{1}{\nn-1}} e^{\frac{G_{\nn}}{(\nn-1)\kt}} \fa^{-\frac{\nn-\mA+1}{\nn-1}}
  \label{eq:Ckt}.
\end{align}

While the above analysis identifies a maximum concentration above which the reaction will become kinetically trapped, we can also identify a minimum concentration below which the median assembly time becomes longer than the maximum timescale of the experiment or simulation $\tmax$
\begin{align}
\cmin \cong \left(\Ahalf / \tmax N k \right)^{-\frac{1}{\nn}} e^{\beta G_{\nn}/\nn} \fa^{-\frac{\nn+1}{\nn}}
\label{eq:cmin}
\end{align}
which applies to both \NA and \YA.
%

A similar analysis can be performed for fixed subunit concentration and varying $\gnuc$, corresponding to changing $p$H or salt concentration. For example, if we assume that pre-critical intermediates with $\nn$ subunits have $\nn-1$ subunit-subunit interactions (Fig.~\ref{fig:scheme}), then $G_\nn=(\nn-1)\gnuc$ and the minimum affinity for assembly in finite time $\tmax$ is shifted according to
\begin{align}
\gbeff = \gb + \frac{\nn+1}{\nn-1} \log \fa
\label{eq:gbeff}.
\end{align}
%
%

\subsection{The influence of conformation dynamics on assembly robustness}
Based on the above scaling estimates, we now examine whether introducing allostery makes assembly  more robust --- \ie, at a finite timescale $\tmax$ relevant to an experiment or a cell, does allostery increase the range of concentrations over which productive assembly occurs?
Specifically, we assume that kinetic trapping precludes productive assembly within $\tmax$, and consider the ratio between the minimum concentration leading to significant nucleation, and the maximum concentration leading to kinetic trapping, $\cratio=\ckt/\cmin$. Since nucleation with allostery must occur within $\tmax$,  we increase the binding affinity according to  Eq.~\eqref{eq:gbeff}, resulting in
\begin{align}
\cratio &\cong \exp(\beta \gbeff/\nn) \fa^{\frac{\mA}{\nn-1}}
\label{eq:cratio}.
\end{align}
Since $\mA=0$ for \YA, allostery with induced fit has no effect on the range of concentrations over which assembly occurs (for moderate interaction strengths), while in the absence of induced fit, allostery renders assembly kinetics more sensitive to concentration (since $\fa<1$). However,  either allostery mechanism  can significantly increase the maximum thermostability of a capsid that can be achieved without kinetic trapping. For a fixed concentration $c_0$, equating Eqs.~\eqref{eq:telong} and \eqref{eq:halftime} shows that the minimum binding free energy $\gkt$ below which kinetic trapping occurs is decreased (higher binding affinity) according to
\begin{align}
\gkt & = \gkt(\fa=1) + \frac{\nn+1-\mA}{\nn-1}\log \fa
\label{eq:gkt}.
\end{align}
The maximum kinetically accessible thermostability of a complete capsid is then controlled by the free energy per subunit, $\gsub=\nb/2\gkt$ with $\nb$ the number of contacts each subunit makes with its neighbors.
For the case we consider below with $\nnuc=5$ and $\nb=4$, $\gsub$ increases by a factor $\nb\mc/\gA/2 \approx 3.33\gA$.  Thus, even a modest activation energy could substantially increase the caspid stability. This effect, together with the asymptotic approach to equilibrium an assembly reaction discussed below, could contribute to observations of unexpectedly large hysteresis between capsid assembly and disassembly.

{\bf Strong allostery  drives robust assembly in the high-affinity limit.}
While the above analysis shows that allostery does not lead to more robust assembly for moderate interaction strengths, we observe dramatically different behavior with strong interactions, $\gb\gg\kt$. It is well-known that unregulated assembly fails due to kinetic trapping in this limit. Since subunit-subunit interactions are effectively irreversible, the effective critical nucleus size is reduced to a dimer regardless of the capsid geometry, and elongation becomes slower than nucleation for any feasible parameters. However, equating Eqs. \eqref{eq:telong} and \eqref{eq:halftime} shows that this trap can be avoided by a sufficiently high $\gA$. We expect trapping to be avoided when the parameter $\ptrap\lesssim1$, with
\begin{align}
\ptrap \cong & \fa^{2-\mA} N^2
\label{eq:ptrap}.
\end{align}
This result holds independent of subunit concentration and capsid geometry, provided the capsid terminates at a finite size and the subunit-subunit affinity is strong enough to stabilize the capsid.  It is important to note that assembly would be sensitive to formation of defective capsids in this limit, which we neglect in this scaling analysis. However, as shown below we do observe this limit in Brownian dynamics simulations where defective assembly is allowed.

\section{Numerical Results}
\label{sec:results}

\subsection{Effect of Allostery on Assembly Robustness}

In this section, we test the scaling predictions against results from the Brownian dynamics (BD) simulations and master equation model.
For the BD simulations, we focus on quantities involving variation of the subunit binding energy parameter $\eb$, as varying this parameter is more computationally tractable than varying subunit concentrations over the range needed to test scaling. Specifically, we test how the assembly yield and median assembly time depend on $\eb$ and the activation energy $\gA$, and we test the prediction that a sufficiently large value of $\gA$ enables assembly even in the limit of high subunit-subunit binding affinity. The latter test is particularly important since the BD simulations allow for the formation of defective capsids. We then  vary concentration using the master equation model. 

\begin{figure*}[hbt]
  \begin{center}
  \includegraphics[width=0.7\columnwidth]{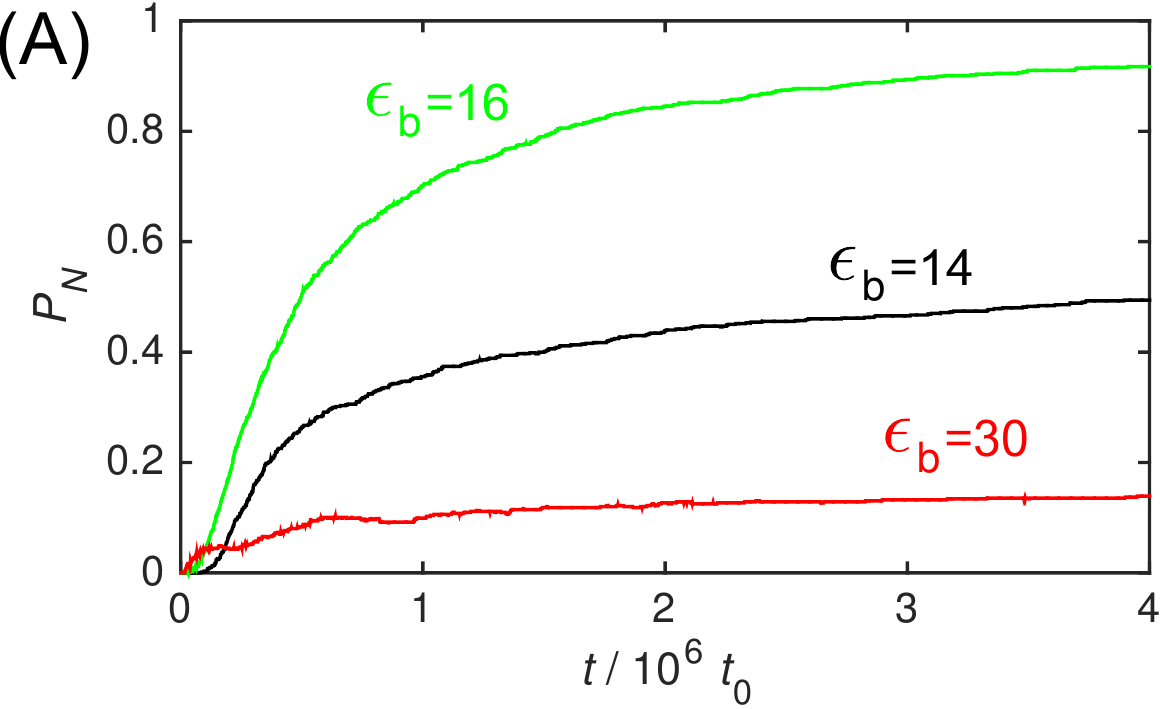}
  \includegraphics[width=0.7\columnwidth]{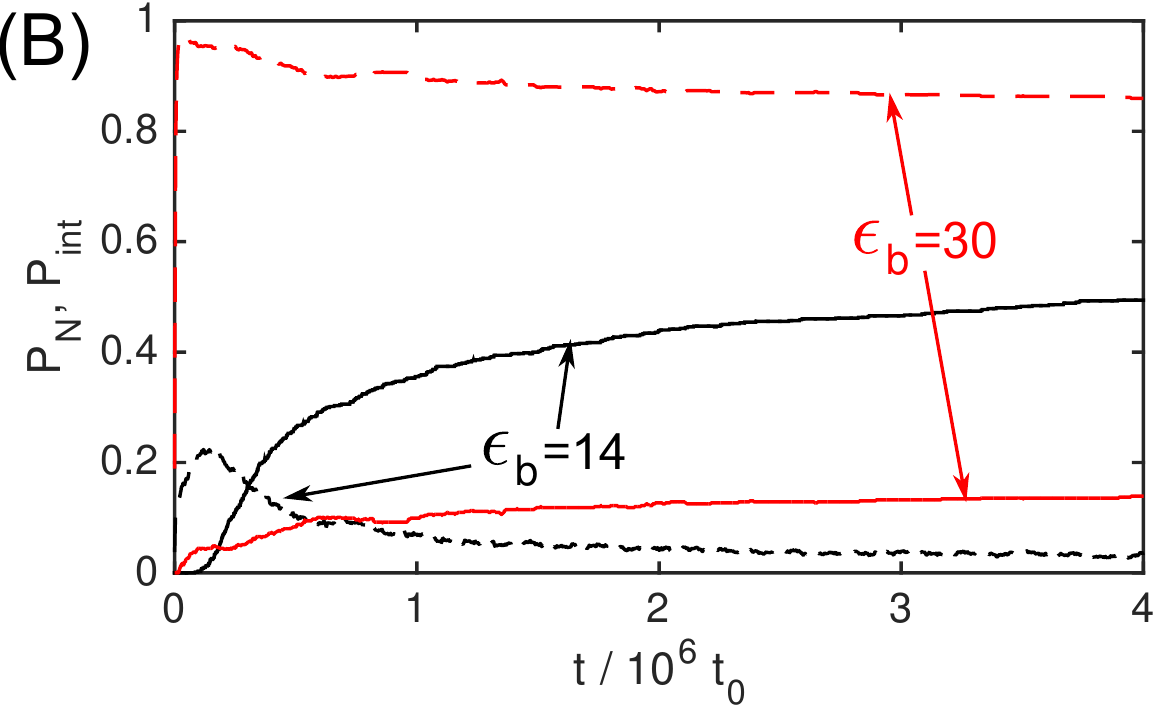}
  \includegraphics[width=0.7\columnwidth]{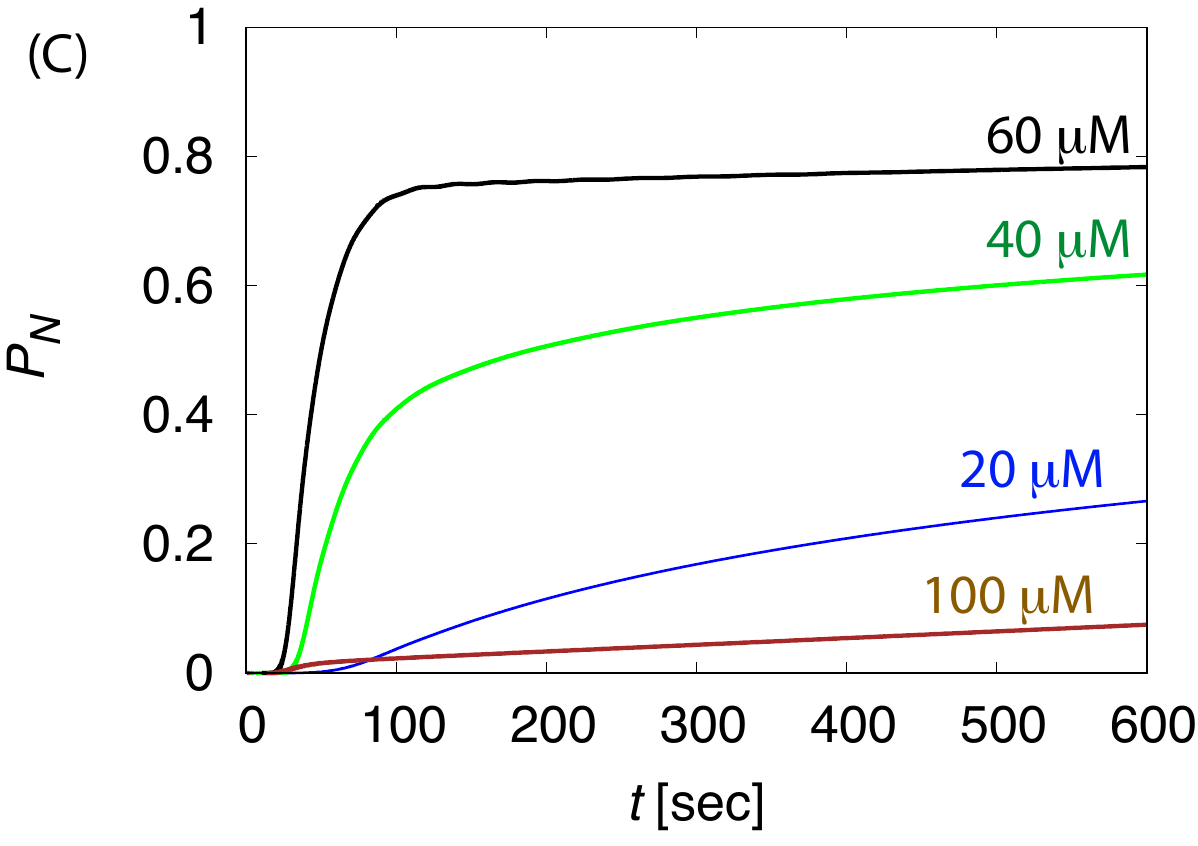}
  \includegraphics[width=0.7\columnwidth]{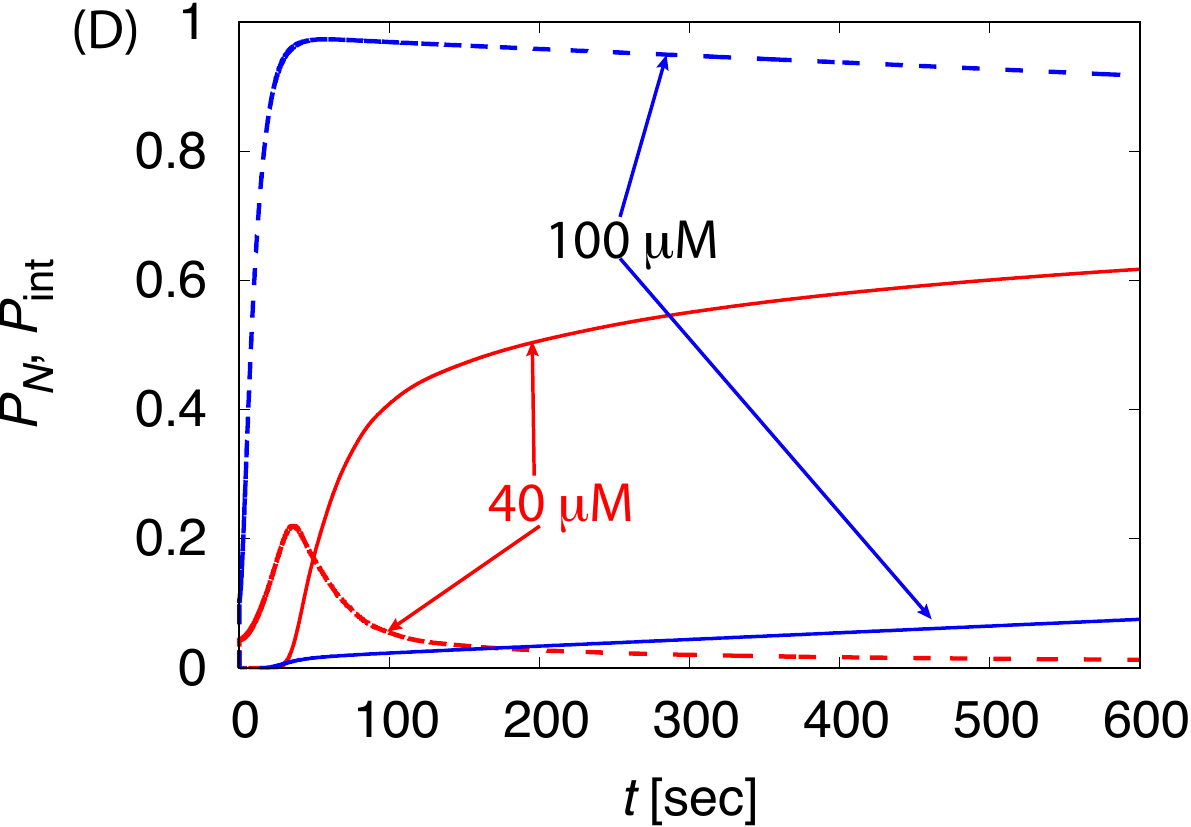}
    \caption{The dependence of assembly kinetics on parameter values for \NC. {\bf (A)} The fraction of subunits in complete capsids $\fc$ observed in Brownian dynamic simulations (BD) is shown as a function of time for indicated values of the binding energy parameter $\eb$. {\bf (B)} The fraction capsid (solid lines) is compared to the fraction of subunits in intermediates $\pint$ (dashed lines) for small and large values of $\eb$. For (A) and (B),  each simulation is run until $1.5\times 10^7 t_0$, and each data point corresponds to an average over 20 independent simulations.  {\bf (C)} The fraction capsid as a function of time measured for the master equation is shown at indicated total subunit concentrations $c_0$, with binding affinity $\gb=-7$ (all energies are in units of $\kt$).  {\bf (D)} The fraction capsid is compared to the intermediate fraction for the master equation, for concentrations below and above the trapping point $\ckt=60\mu$M (Eq.~\ref{eq:Ckt}).}
  \label{fig:fcVsTime}
  \end{center}
\end{figure*}

We begin by examining the time dependence of the fraction of subunits in complete capsids ($\fc$) as the binding energy is varied. A capsid is defined as a structure containing 20 subunits, each with strong interactions with three neighbors. This definition counts only configurations which correspond to small fluctuations around the icosahedral ground state.
We note that experimental measurements of assembly kinetics commonly employ SEC, which monitors $\fc$, or light scattering, which under certain conditions monitors the mass-averaged molecular weight of assemblies \cite{Zlotnick1999}. Ref.~\cite{Hagan2010} showed that $\fc$ and assembly molecular weight closely track each other below the crossover concentration $\cc$ (Eq.~\eqref{eq:cc}), and quantities such as the lag phase and nucleation time follow the same scaling laws when calculated from either observable.  Therefore, here we present only $\fc$.

Fig.~\ref{fig:fcVsTime}A shows $\fc$ as a function of time for several values of $\eb$ with the \NC model.
We see that initially as $\eb$ increases the lag phase shortens, $\fc$ rises in time more rapidly, and asymptotes at a higher value. However, at the highest $\eb$ shown, $\eb = 30$, the fraction capsid quickly saturates at a low value, indicative of kinetic trapping. To illustrate this point, Fig.~\ref{fig:fcVsTime}B shows $\fc(t)$ for two values of $\eb$, overlaid with a plot of the fraction of subunits in intermediates, $\pintt(t) = \sum_{n=2}^{N-1} n c_n(t)/c_0$.
 We see that for $\eb=14$ the intermediates peak near the end of the lag phase, and then rapidly fall as capsids are produced. In contrast, for $\eb=30$, nearly all subunits are trapped in intermediates. Thus capsids are only slowly produced when larger intermediates scavenge subunits from smaller ones. Since we find that it correlates well with other measures of kinetic trapping, we use $\pint$ to characterize trapping as a function of parameter values.

For comparison, Figs.~\ref{fig:fcVsTime}C,D show the time-dependence of $\fc$ and $\pint$ calculated from the master equation with varying subunit concentration \cite{PintNote}.  We see that the  kinetics and the relationship between $\pint$ and the onset of kinetic trapping are consistent with the BD results.

\begin{figure}
\centering
  \includegraphics[width=0.7\columnwidth]{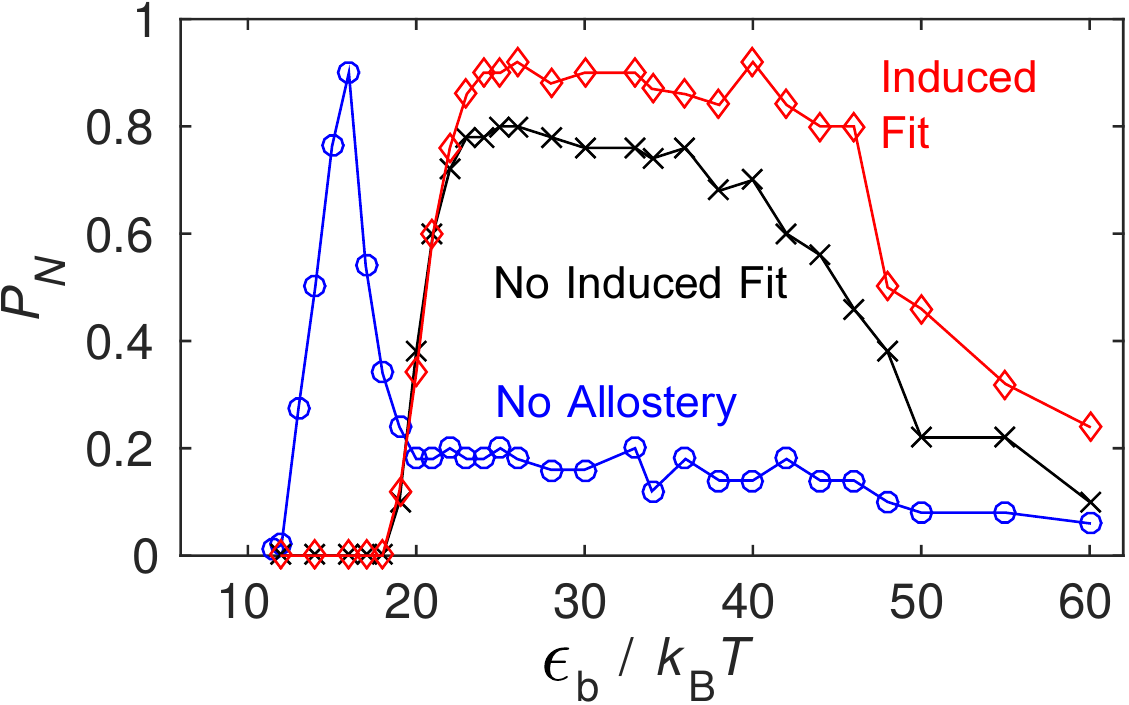}
  \includegraphics[width=0.7\columnwidth]{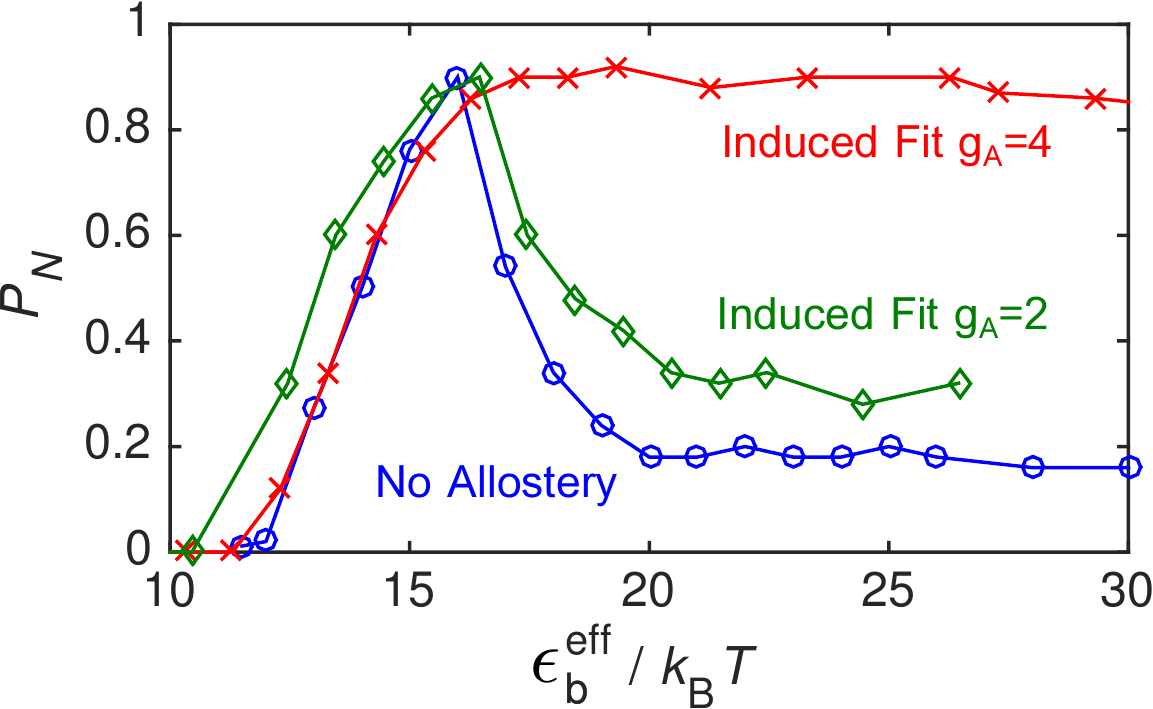}
\caption{ (top) Capsid fraction $\fc$ as a function of the binding energy parameter $\eb$ in Brownian dynamics (BD) simulations for the three conformation cases, with $\gA=4\kt$.  (bottom) The capsid fraction is shown as  a function of the shifted binding energy $\ebeff$ (Eq.~\ref{eq:ebeff}) for \NC and \YA with two activation energy values. For both panels, other parameters are as in Fig.~\ref{fig:fcVsTime}. }
  \label{fig:BDcapsid_fraction}
\end{figure}

Next, we consider how assembly robustness changes when allostery is introduced. Fig.~\ref{fig:BDcapsid_fraction}A shows $\fc$ measured at long but finite times as a function of $\eb$ for the three conformation cases with $\gA=4$. We see that in both cases, allostery shifts the onset of assembly to dramatically higher binding affinity. More significantly,  assembly stays remarkably productive up to extremely high values of $\eb$ for both allostery mechanisms. This observation corresponds to the high affinity limit discussed above (Eq.~\eqref{eq:ptrap}). Analysis of simulation trajectories shows that the eventual decline in $\fc$ at high $\eb$ arises due to the formation of defective capsids.

To show how assembly robustness depends on the strength of allostery, we compare $\fc$ calculated for the \NC with \YA at two values of $\gA$ in Fig.~\ref{fig:BDcapsid_fraction}B. To aid in comparing these cases, we have `undone' the shift in the binding energies suggested by Eq.~\eqref{eq:gbeff} according to
\begin{align}
\ebeff=\eb + \frac{\nn+1}{\nn-1} \log \fa
\label{eq:ebeff},
\end{align}
so that the onset of productive assembly occurs roughly at the same value of $\ebeff$ for each case.
We note that the curves do not line up perfectly because this mapping is only approximate, since the free energy includes binding entropy factors whose magnitude depend logarithmically on $\eb$ (see Eq.~\eqref{eq:sb} in the appendix) and the degeneracy of available binding sites on a given structure \cite{Hagan2006,Hagan2009,Hagan2011}.
We see that the smaller activation energy $\gA=2$ leads to assembly over approximately the same range of $\ebeff$ as for the \NC case, supporting our earlier conclusion that, below the high affinity limit, allostery does not significantly increase assembly robustness.

\begin{figure}
\centering
    \includegraphics[width=0.7\columnwidth]{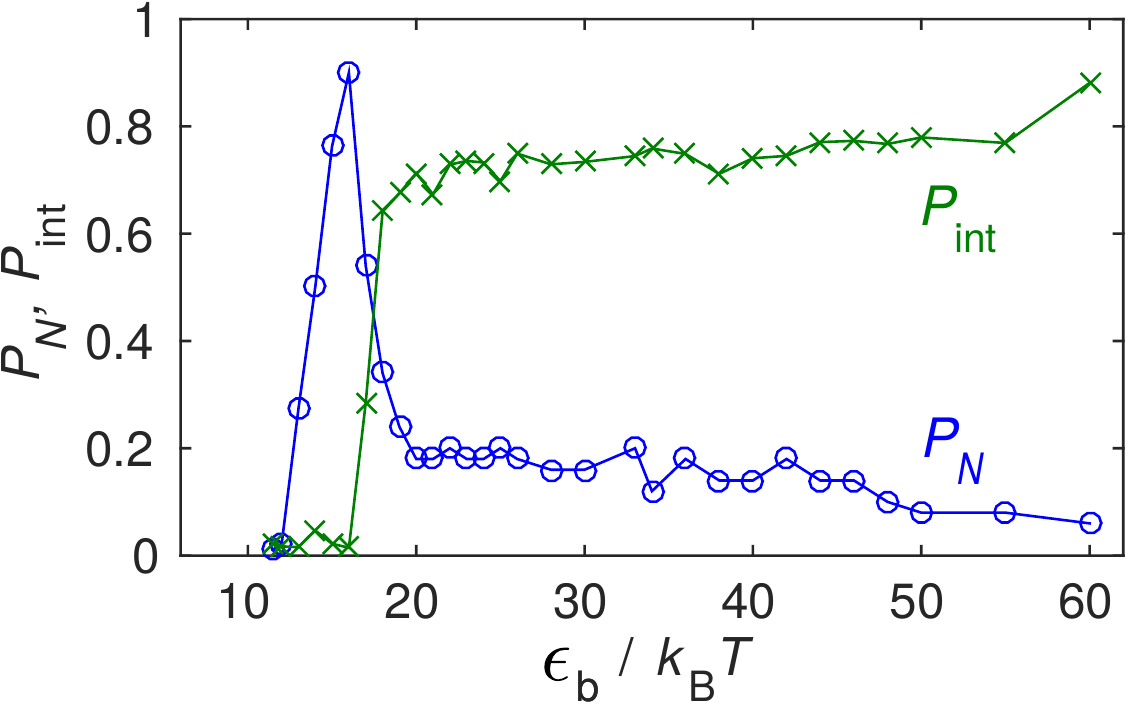}
\caption{ Capsid fraction $\fc$ and intermediate fraction $\pint$ as a function of the binding energy $\eb$ in Brownian dynamics (BD) simulations for \NC.  }
  \label{fig:Pint}
\end{figure}

To further understand the onset of the high affinity limit and robust assembly, we examine the dependence of kinetic trapping on parameter values. First, to support the earlier statement that $\pint$ is a good metric for the extent of kinetic trapping, Fig.~\ref{fig:Pint} compares $\fc$ and the intermediate fraction $\pint$ measured at long times for \NC;  we see that a rise in $\pint$ correlates with the decline in $\fc$. 

\begin{figure}
\centering
  \includegraphics[width=0.7\columnwidth]{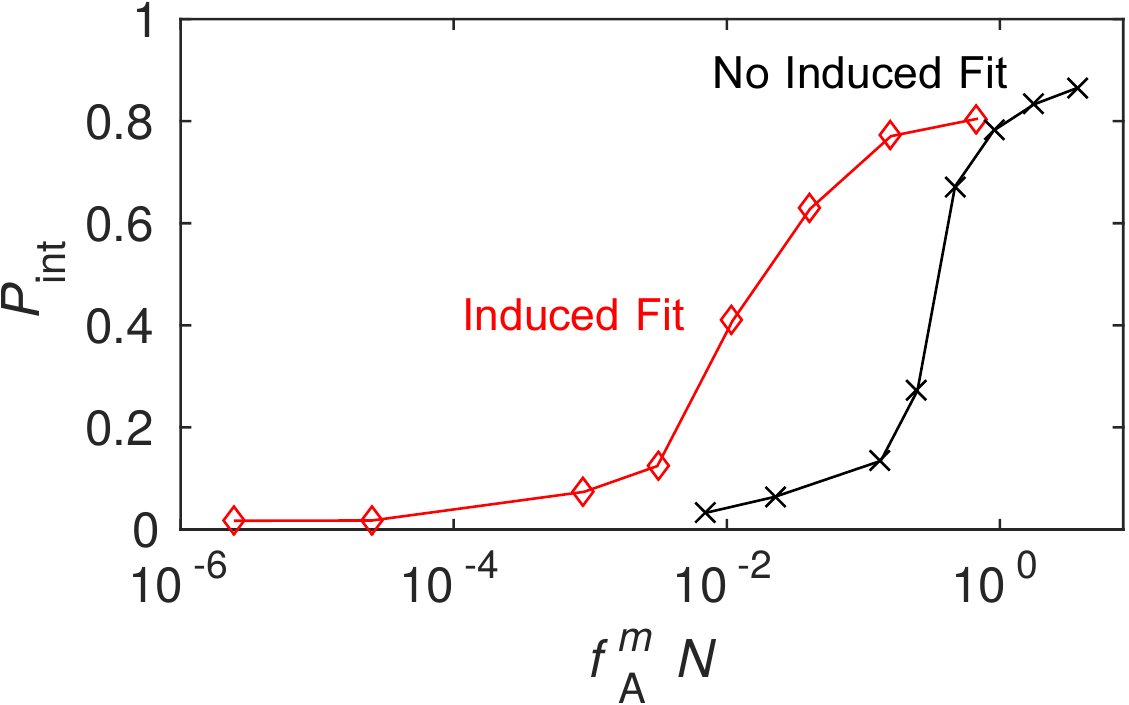}
  \includegraphics[width=0.7\columnwidth]{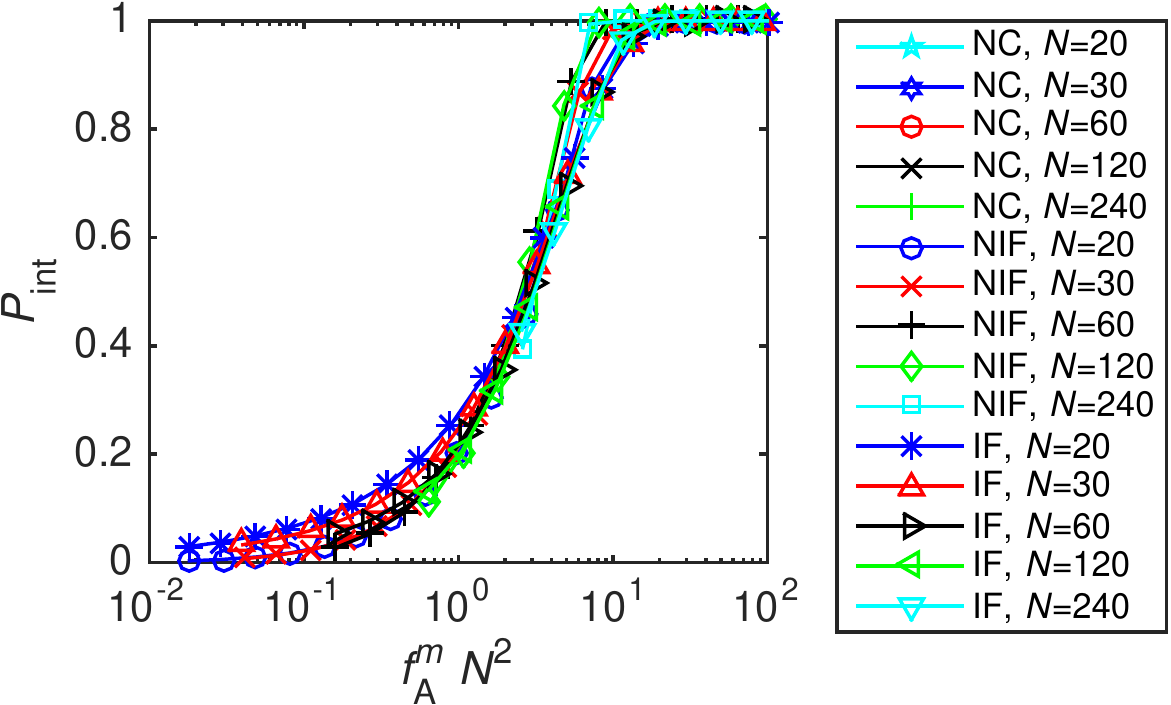}
\caption{{\bf(top)} Kinetic trapping in BD simulations. The magnitude of trapping $\pint$ measured in BD simulations is shown
 in the limit of very high subunit binding strength ($\eb=40$).  Results are shown as a function of the high affinity trapping parameter $\ptrap=\fa^m N^2$, with $m=1$ or $2$ respectively for the \NA and \YA cases (see Eq.~\eqref{eq:ptrap}) and $N=20$. The parameter $\fa$ was tuned by varying $\gA/\kt$ from 1 to 6. Other parameters are as in Fig.~\ref{fig:fcVsTime}. {\bf (bottom)} Relationship between capsid size and sensitivity to trapping in the high-affinity limit from the master equation. The magnitude of trapping $\pint$ calculated for $\gb=-25$ is shown for the three conformation cases and capsid sizes $N\in[20,240]$.  The parameter $\fa$ was tuned by varying $\gA/\kt$ from 0 to 10. }
  \label{fig:BDVaryGactive}
\end{figure}
 
Next we focus on high binding affinity, $\eb=30$, and show $\pint$ in Fig.~\ref{fig:BDVaryGactive}A for different values of $\gA$, plotted against the trapping parameter $\ptrap$ (Eq.~\eqref{eq:ptrap}).
 We see that for \NA the onset of trapping occurs when $\ptrap$ is of order one as expected, but trapping for \YA is shifted to lower values of $\ptrap$.
Analysis of simulation trajectories suggests that this difference arises due to the formation of malformed capsids, which are are more prevalent in the \YA trajectories. We can understand this observation by noting that the probability of a defective capsid increases with the rate of subunit addition, which is larger for \YA (since inactive subunits can bind to growing capsids, and because a given value of $\ptrap$ corresponds to higher $\fa$ for \YA in comparison to \NA). As further evidence for the importance of defective capsids, Fig.~\ref{fig:BDVaryGactive}B shows the equivalent plot from the rate equation model, which does not allow for malformed capsids. In that case, we see that results from both allostery mechanisms fall approximately on the same curve. We have also tested the predicted scaling against capsid size $N$ in that figure by including results from rate equation calculations with $N$ varying from 20 to 240.

\begin{figure}[hbt]
  \begin{center}
  \includegraphics[width=0.7\columnwidth]{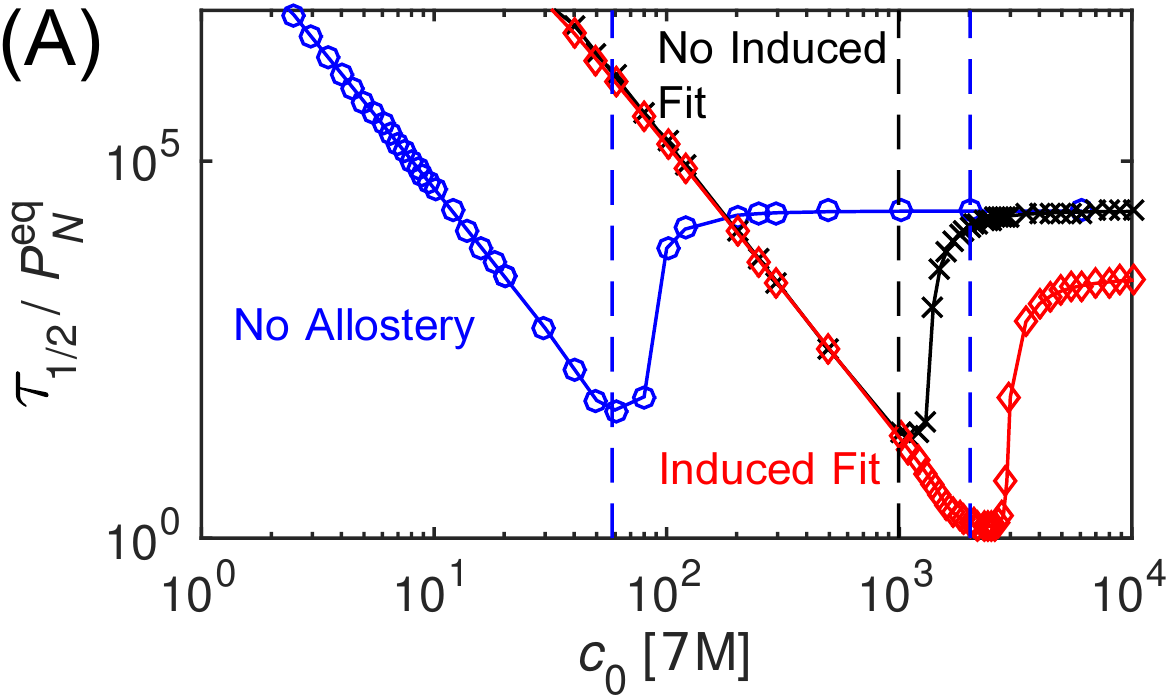}
  \includegraphics[width=0.7\columnwidth]{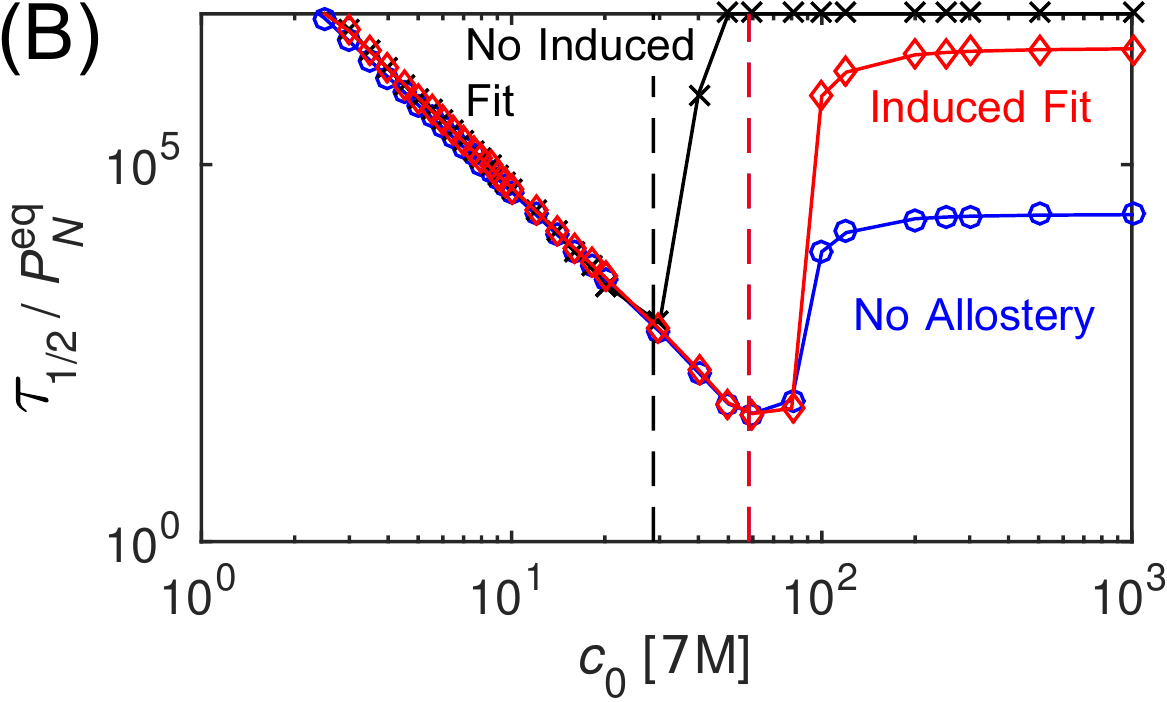}
 \includegraphics[width=0.7\columnwidth]{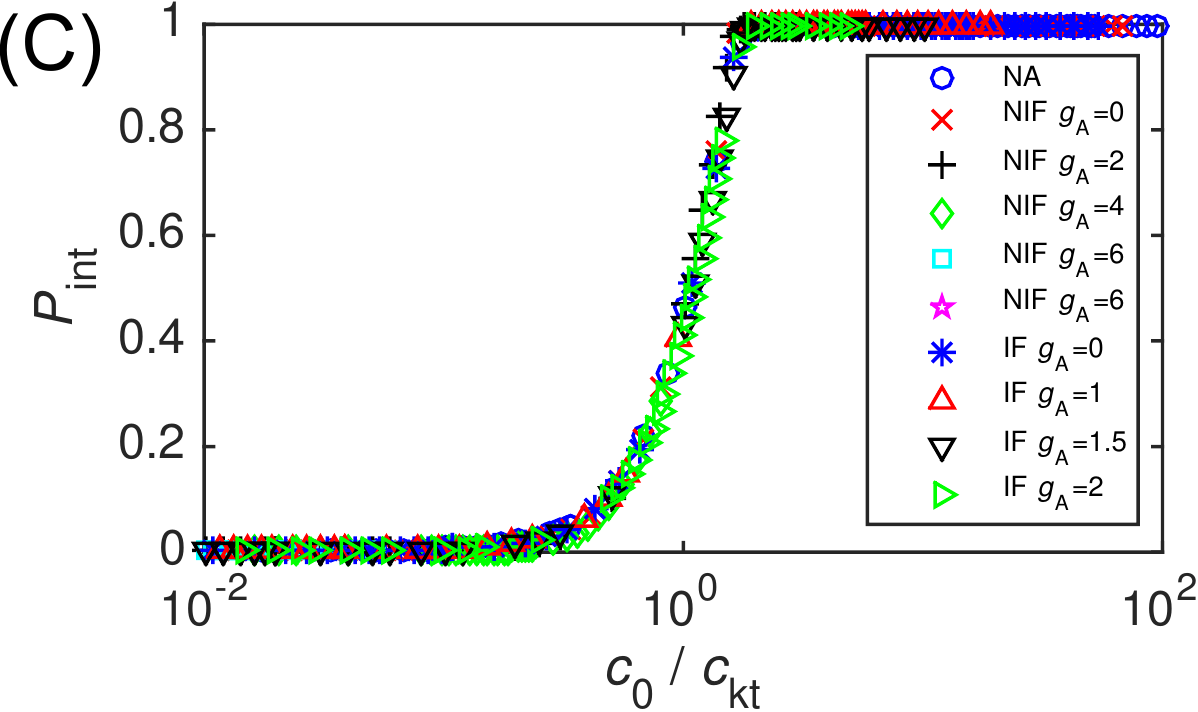}
    \caption{The effect of allostery on assembly timescales calculated from the master equation. {\bf (A)} The median assembly time $\halftime$ is shown as a function of initial subunit concentration for the three allostery cases, with the subunit binding affinity $\gb=-7$ and the activation energy $\gA=-2$.  For each case, the vertical dashed line indicates the point of kinetic trapping $\ckt$ calculated by Eq.~\eqref{eq:Ckt}.   {\bf (B)} The median assembly time is shown as a function of concentration for the three cases, but with the subunit binding affinity shifted according to  Eq.~\eqref{eq:gbeff}, $\gb=-7$ for NC and $\gb=-10.54$ for NA and A, with $\gA=2$. The vertical lines indicate the predicted location of $\ckt$ for these modified subunit affinities. {\bf (C)} The maximum fraction of subunits found in intermediates, $\pint$, is shown as a function of initial subunit concentration normalized by the trapping threshold $\ckt$ (Eq.~\ref{eq:Ckt}), for indicated allostery cases and values of the activation energy $\gA$. }
  \label{fig:halftime}
  \end{center}
\end{figure}
\begin{figure}
\centering
  \includegraphics[width=0.7\columnwidth]{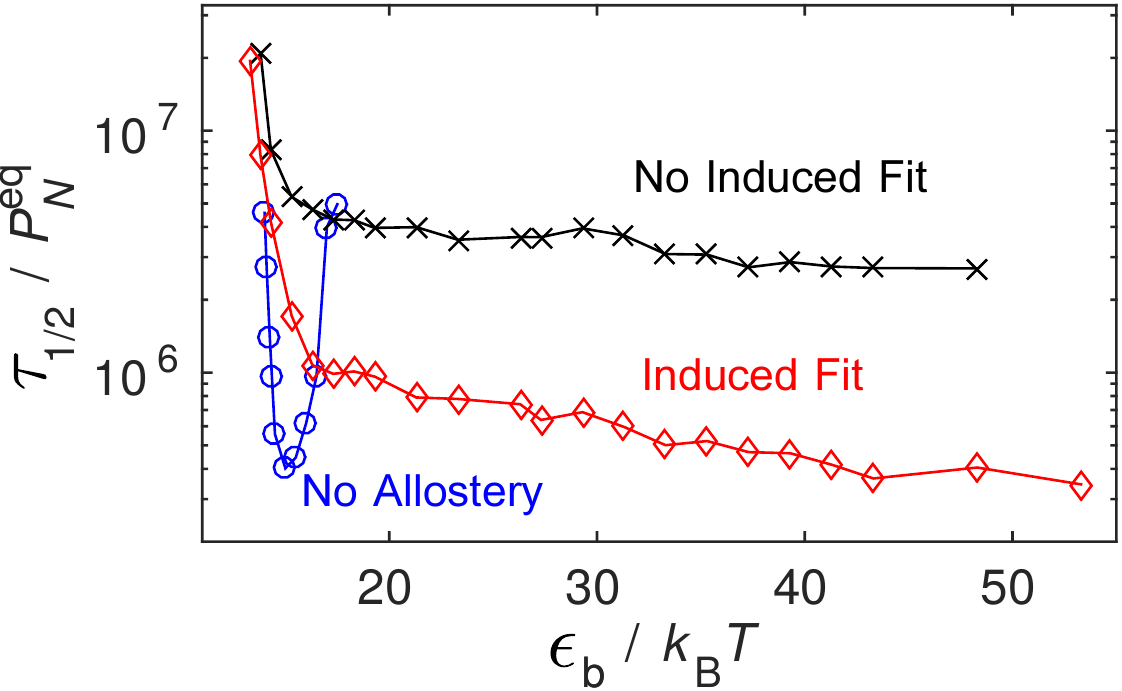}
\caption{ The median assembly time $\halftime$ measured in BD simulations is shown for the three conformation cases as a function of $\ebeff$, with $\gA=4$. Other parameters are as in Fig.~\ref{fig:fcVsTime} except simulations were run until $2.6\times10^7 t_0$.
 }
  \label{fig:BDmeanTime}
\end{figure}

Finally, we consider the effect of allostery on assembly timescales. Fig.~ ~\ref{fig:halftime}A shows the median assembly time $\halftime$ (defined by $\fc(\halftime)=1/2$) calculated from the master equation as a function of initial subunit concentration for each conformation case, with modest interaction parameter values ($\gb=-7$, $\gA=2$).  This plot allows testing of several scaling predictions. Firstly, we see that both in the presence and absence of allostery, the $\halftime$ at low concentrations scales as $c_0^{\nnuc-1}$ as predicted by Eq.~\eqref{eq:halftime}.  As expected, median times are equal for \NA and \YA at low concentrations where the reaction is nucleation-limited, since the autostery does not affect nucleation in our model. Secondly, the dashed vertical lines show the point of kinetic trapping $\ckt$ predicted by Eq.~\eqref{eq:Ckt} for each case. We see that the median assembly time takes off above these threshold concentrations, indicating the presence of trapping.

Notice that even a modest conformation energy $\gA=2$ shifts the region of productive assembly to unrealistically high subunit concentrations, due to the scaling of nucleation times with $\fa$. However, this effect can be counteracted by increasing the subunit binding affinity. To illustrate this point, and to test the predicted effect of conformation specificity on assembly robustness (Eq.~\eqref{eq:cratio}), we also calculated median assembly times with the binding affinity for the allostery cases increased according to Eq.~\eqref{eq:gbeff} (Fig.~\ref{fig:halftime}B).  As anticipated,  the nucleation times become equal at low concentrations, and the \YA and \NC cases enjoy the same range of productive assembly, while the \NA case becomes kinetically trapped at a lower concentration due to its increased elongation time relative to nucleation. To further test the expected scaling with activation energy, we show the trapping measure $\pint$ as a function of subunit concentration normalized by $\ckt$ for all three cases over a range of $\gA$ in Fig.~\ref{fig:halftime}C. We see that in all cases trapping takes off as the threshold concentration is crossed. 

Fig.~\ref{fig:BDmeanTime} shows the variation of $\halftime$ with the shifted binding energy for the three allostery cases calculated from BD simulations. We see that the nucleation-dominated regime, marked by an exponential decrease in $\halftime$ with increasing $\ebeff$, roughly overlaps for the three cases, suggesting that the mapping is approximately correct. Notice that with allostery, $\halftime$ becomes constant at large $\ebeff$; this corresponds to the high affinity limit.

\subsection{Implications of allostery for parameter estimations from experimental data}
\begin{figure}[hbt]
  \begin{center}
  \includegraphics[width=0.7\columnwidth]{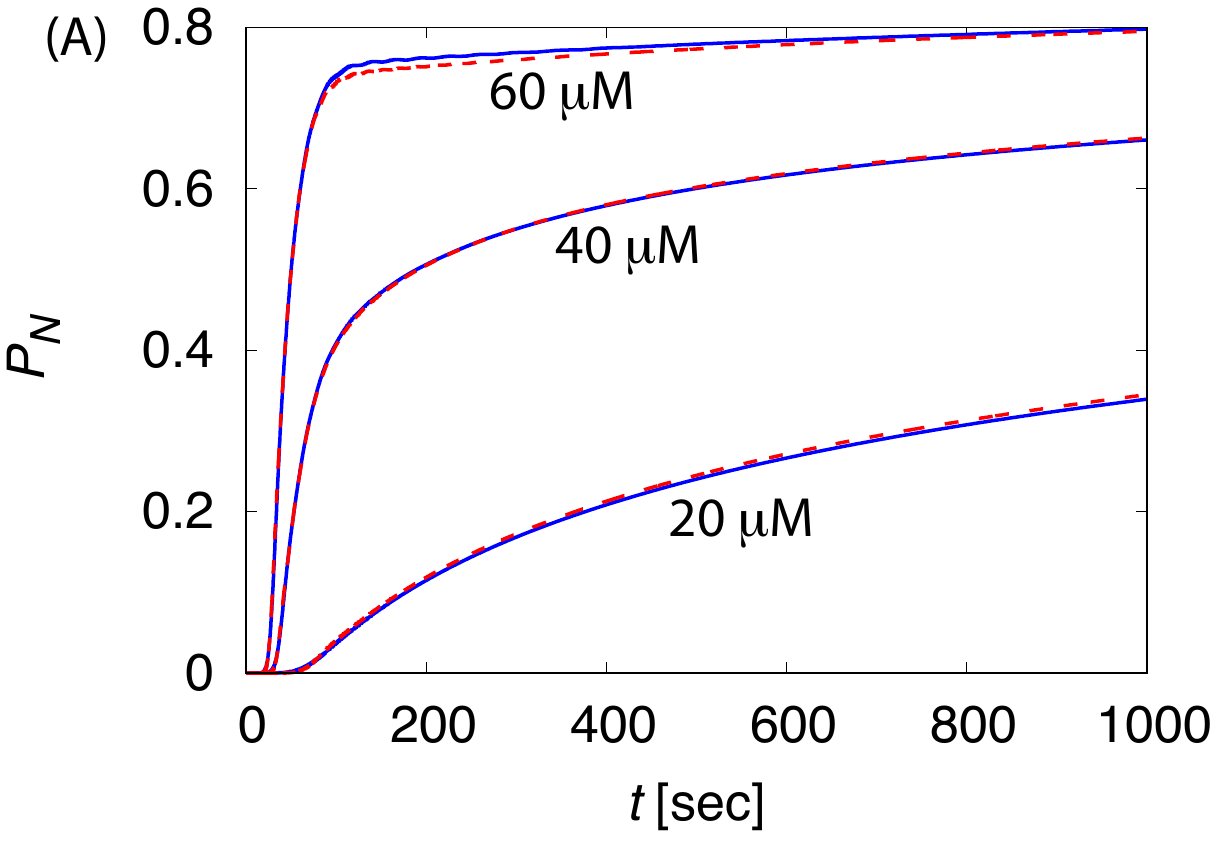}
  \includegraphics[width=0.7\columnwidth]{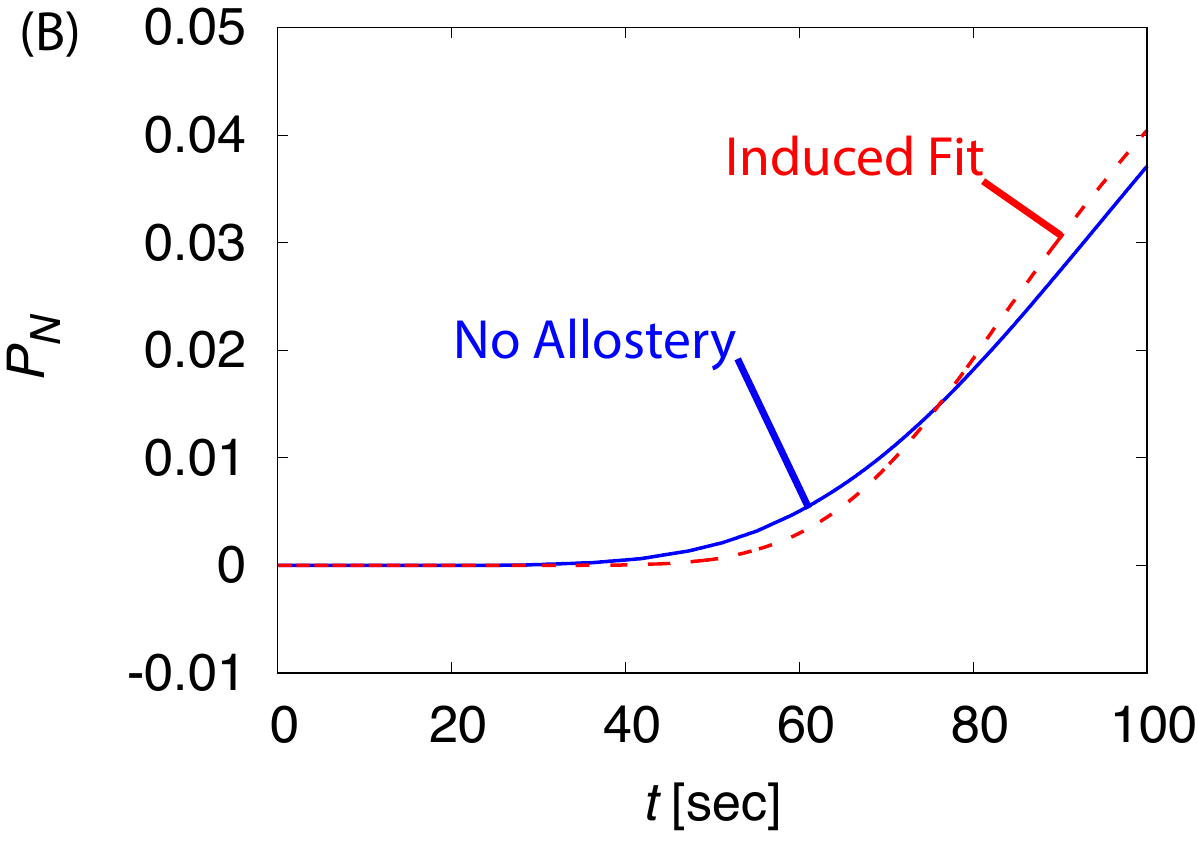}
    \caption{The time-dependence of assembly is compared between the \NC (solid lines) and \YA (dashed lines) cases for $c_0=20\mu$M.  {\bf (A)} The fraction capsid is shown for indicated initial subunit concentrations.    {\bf (B)} The early time course of assembly is shown for both cases at $c_0=20\mu$M, showing that \YA exhibits a sharper take off at the end of the lag phase in comparison to \NC.}
  \label{fig:fcVsTimeCompare}
  \end{center}
\end{figure}
In this section we compare assembly kinetics with and without allostery, and consider ramifications for parameter estimations made using models that do not account for allostery.
Fig.~\ref{fig:fcVsTimeCompare} compares the time-dependence of $\fc$ at several concentrations for the \NC and \YA cases calculated from the master equation, with the subunit binding affinity in the \YA case adjusted according to Eq.~\eqref{eq:gbeff}. In particular, with $\gb=-7$ for \NC, we obtain $\gb=-10.54$ for \YA with $\gA=2$.
We see that this relationship, derived to relate assembly kinetics in the nucleation-dominated regime, leads to assembly kinetics which also closely match throughout early times.

The most significant difference between the kinetics with and without allostery appears at a very early times. As the kinetics transition from the lag phase to the rapid production of capsids, the takeoff is more rapid for the \YA case. A similar observation was made by Chen \etal \cite{Chen2008} and shown to be consistent with light scattering from assembling BMV viruses. This observation can be understood by noting that, due to the stronger binding affinity in the \YA case, kinetics in the elongation phase are closer to irreversible and thus the distribution of lag times is more sharply peaked.

Despite its modest impact on the form of kinetics, allostery can dramatically skew quantitative parameter values estimated from experimental data. Fig.~\ref{fig:fcVsTimeCompare} demonstrates that a fit against assembly kinetics not accounting for allostery would underestimate $\gb$, by an amount proportional to $\gA$ (see Eq.~\eqref{eq:gbeff}).  We emphasize that the shift factor in $\gbeff$ is independent of subunit concentration. Thus, testing data fits against multiple subunit concentrations does not necessarily identify the presence of allostery. A binding affinity underestimated at one subunit concentration (due to not accounting for allostery) would be consistent with data at other subunit concentrations.
 \begin{figure}[hbt]
  \begin{center}
  \includegraphics[width=.99\columnwidth]{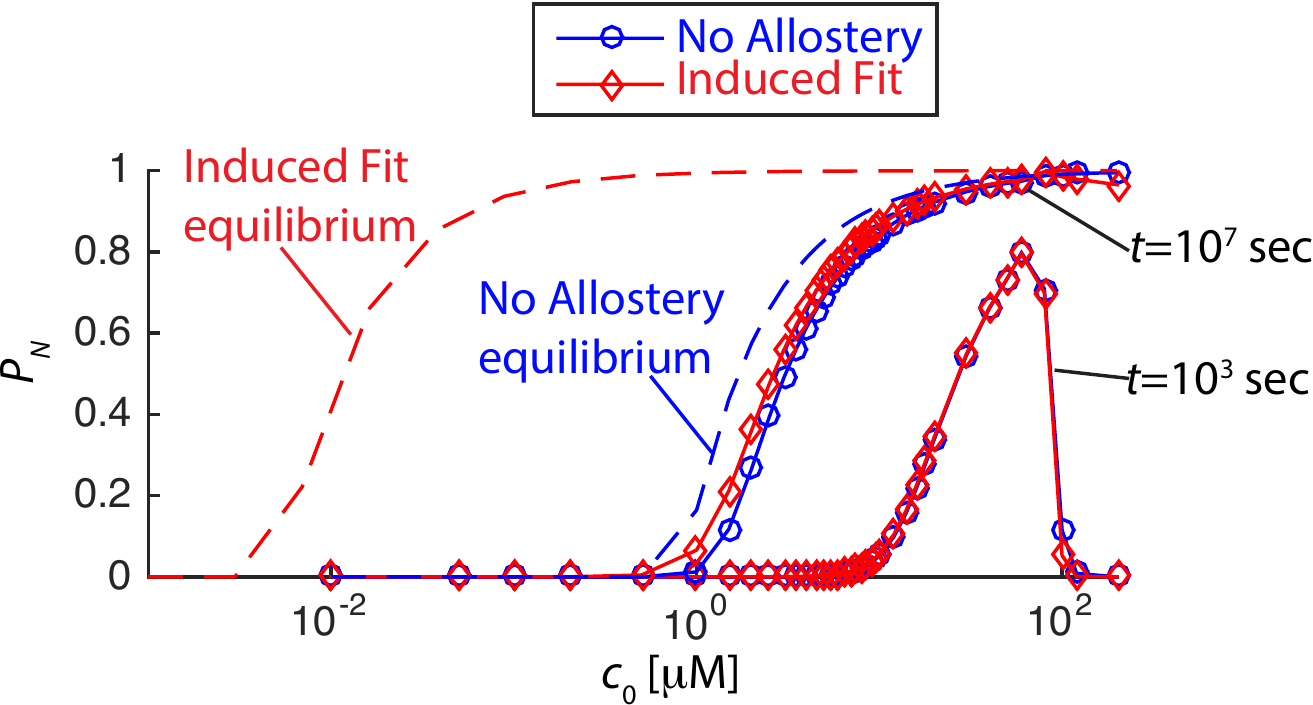}
    \caption{The difficulties of detecting allostery from finite-time experiments. The fraction of subunits in capsids calculated from the master equation as a function of total subunit concentration is shown for $10^3$ and $10^7$ seconds, for \NC (\textcolor{blue}{o} symbols) and \YA (\textcolor{red}{$\diamond$} symbols). The equilibrium values of $\fc$ are shown for each case as dashed lines.   Parameters are $\gb=-7$ for \NC and $\gb=-10.54$ and $\gA=2$ for \NC.
}
  \label{fig:FinalFc}
  \end{center}
\end{figure}

A second approach to estimate subunit binding affinities is to measure the ratio of capsid to free subunits at very long times, and fit the results to the equilibrium law of mass action \cite{Ceres2002}.
As has been pointed out previously \cite{Roos2010,Hagan2010,Hagan2014}, $\fc$ only asymptotically approaches its equilibrium value, and thus fits assuming equilibrium will tend to underestimate $\gb$. We find that unlooked for allostery can dramatically enhance underestimation, since the lack of distinguishability seen in Fig.~\ref{fig:fcVsTimeCompare} persists for very long times.
In Fig.~\ref{fig:FinalFc} we show the fraction capsid $\fc$ as a function of subunit concentration predicted from the master equation with and without allostery, for the same interaction parameters as in Fig.~\ref{fig:fcVsTimeCompare}.
We see that the curves overlap perfectly at 1000 seconds, and still nearly overlap at $10^7$ seconds (about 4 months). By this time, the \NC system has nearly reached its equilibrium, and thus the equilibrium subunit binding affinity estimated from these results would be reasonably accurate. However, for \YA the binding affinity is underestimated by $3.5\kt$. Larger values of $\gA$ lead to more severe underestimation with similar lack of distinguishability between the two cases.

\section{Discussion and Outlook}
\label{sec:discussion}
It has been well established that the assembly of rigid subunits into ordered structures requires weak, reversible interactions, as interactions which are strong in comparison to the thermal energy lead to kinetic traps \cite{Whitelam2015,Hagan2014,Zlotnick2003,Hagan2006,Rapaport2008}.
Here, we have used computational and theoretical models to investigate how this constraint changes when the subunits have internal degrees of freedom, such as conformational states, allowing them to change their capacity for interaction during the assembly process.

We find that a sufficient bias in the free subunit population toward the inactive conformation allows productive assembly at very high subunit concentrations or binding affinities. In particular, allostery differentially regulates the rates of nucleation and elongation, thus suppressing the kinetic trap that arises when free subunits are depleted before capsid elongation is completed.   To our knowledge, such robust assembly has not observed in vitro, but this effect could be important for reactions such as the assembly of the mature HIV capsid which occurs at high concentration within the budded viral particle \cite{Sundquist2012}. However, this mechanism does not provide complete protection against kinetic traps that arise due to defective capsid assembly, as we observed in the Brownian dynamics simulations at very high $\eb$. The prevalence of such defects, and hence the actual range available for productive assembly, will depend on the orientational specificity of the subunit-subunit interactions \cite{Hagan2006,Hagan2011}, which has not yet been evaluated for specific capsid proteins.

Given that increasing $\gA$ can qualitatively change assembly robustness, it would be of interest to measure the subunit conformational equilibrium as a function of protein sequence and solution conditions.  We have recently used all-atom simulations to estimate a free energy difference of about $3\kt$ between two quasi-equivalent conformations (meaning different conformations found at positions with different local symmetry in the capsid) of the MS2 coat protein. However, we note that the accuracy of such calculations is necessarily limited by force field accuracy and the quality of sampling. The simulations also found that RNA binding could significantly shifts the populations. While similar calculations are possible for the active/inactive transition, they will be complicated by the fact that the inactive `conformation' can be an ensemble with significant structural diversity \cite{Packianathan2010}.

 Our findings have several implications for interpreting mechanisms and estimating interaction parameters from experimental data.
Firstly, we find that the kinetics of an assembly reaction with allostery are quite difficult to distinguish from those of a reaction with no conformation dependence at moderate parameters (Fig.~\ref{fig:fcVsTimeCompare}).
 One commonly used test for the quality of a model is to fit interaction parameters to kinetics at one subunit concentration, and then test their predictions against kinetics measured at other concentrations. However, the binding affinity adjustment Eq.~\eqref{eq:gbeff} used to match parameters between the \NC and \YA cases in Fig.~\ref{fig:fcVsTimeCompare} is independent of concentration, and we find that the kinetics match quite closely even into the regime where kinetic trapping starts to set in. Thus, a binding affinity underestimated at one subunit concentration (due to not accounting for allostery) would also appear consistent with data at other subunit concentrations.
While it might be argued that such a close match arises because of the simplicity of the rate equation model, parameter estimation from experimental data generally relies on models with similar levels of approximations to enable computational tractability. Moreover, we observe a similar matching of kinetics in the BD simulations. Thus, our results suggest that strong emphasis should be placed on matching the very early phases of assembly kinetics during parameter estimation.  The importance of fitting the lag phase was also suggested by Chen \etal \cite{Chen2008}, who furthermore demonstrated this phase can be monitored by light scattering with millisecond resolution. Techniques sensitive to individual capsids (\eg \cite{Zhou2011}) will allow further investigation of early-time kinetics.

As we note above, by shifting productive assembly to higher binding affinities, allostery increases the maximum capsid thermostability that is kinetically accessible. Thus, allostery, in the form of conformational transitions during assembly, may offer an alternative strategy to post-assembly conformational changes or covalent modifications used by some bacteriophages to stabilize their capsids (\eg \cite{May2012a,Veesler2012}).

Finally, there are several effects we did not investigate here which could lead to additional control over assembly. We have focused on the limit in which subunit conformational transitions are fast in comparison to assembly timescales. Our approaches are easily generalized to other conformational timescales. A preliminary investigation showed that decreasing the conformational transition rate allows higher assembly yields in the high affinity limit for the \YA case (similar to the case of bacterial flagella assembly \cite{Asakura1968}). While we focus here on allostery at the level of protein-protein interactions, there are many examples in the literature suggesting that interaction with non-protein components, such as RNA, lipid membranes, or small molecule assembly effectors can exert additional allosteric control on protein conformations (see \cite{Perlmutter2015}). Understanding how these multiple regulatory mechanisms cooperate to control the time, place, and rate of assembly will lead to a more complete understanding of viral life cycles, and also may identify new strategies for designing human-made assembly systems.

\section{Appendix}
\subsection{Comparison between Computational and Master Equation Models}
\label{sec:compare}
 \begin{figure}[hbt]
  \begin{center}
  \includegraphics[width=.7\columnwidth]{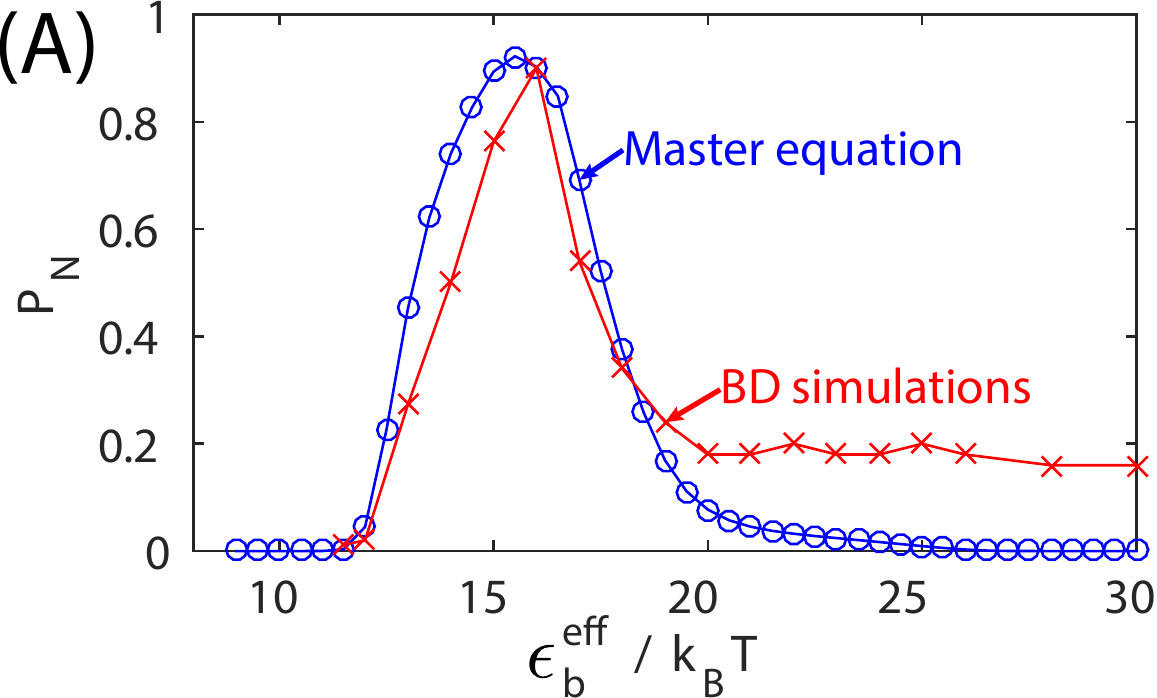}
  \includegraphics[width=.7\columnwidth]{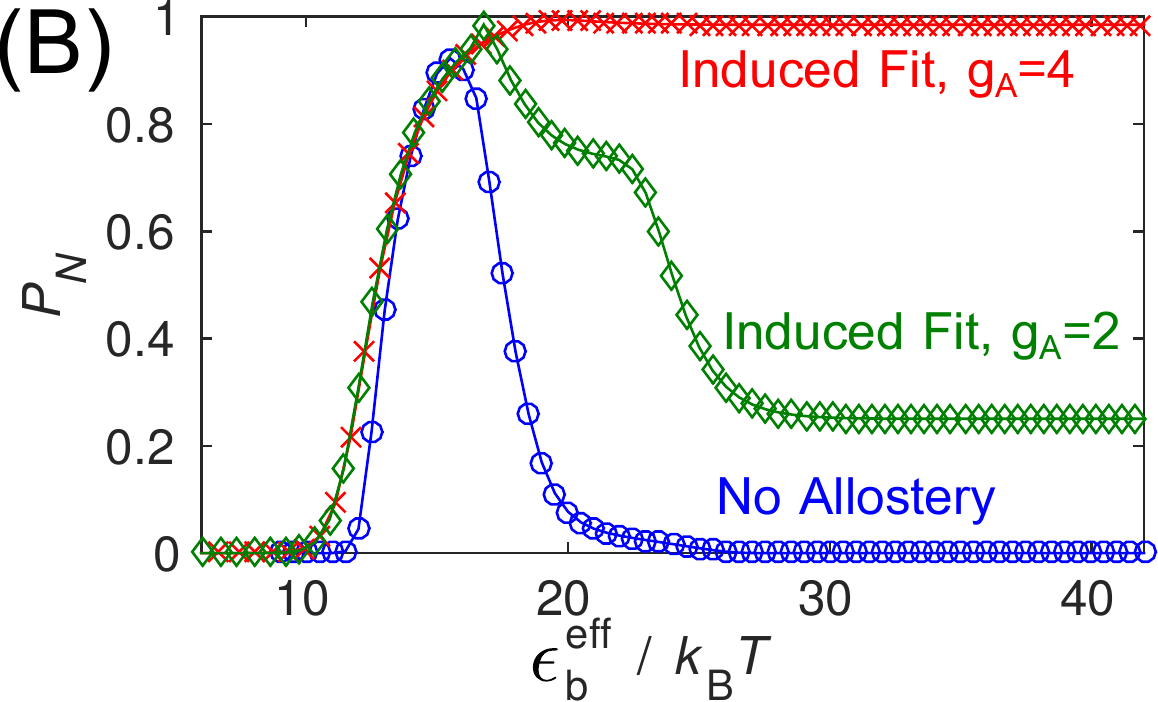}
   \caption{{\bf (A)} The fraction capsid $\fc$ predicted by BD simulations and the master equation as a function of the binding energy parameter $\eb$ for \NC, for $t=10^7t_0$. Parameters for the BD simulations are as in Fig.~\ref{fig:fcVsTime}, and the master equation parameters are set correspondingly (see the text in Appendix ~\ref{sec:compare}).  {\bf (B)} The master equation results for fraction capsid are shown as  a function of the shifted binding energy $\ebeff$ (Eq.~\ref{eq:gbeff}) for \NC and \YA with two activation energy values (compare to the bottom panel of Fig.~\ref{fig:BDcapsid_fraction}).  The only adjustable parameter for the master equation in these results is the association rate constant $k$, which was set (by eye) to $k=0.003 \sigma^3/t_0$.
}
  \label{fig:compare}
  \end{center}
\end{figure}
In the main text our analysis of the master equation focuses on capsid sizes and parameters consistent with experiments on HBV \cite{Zlotnick1999}.  Here we consider parameters roughly corresponding to those of the BD simulations to further evaluate the degree of similarity between master equation and BD results. To minimize reliance on data fitting, we calculated the subunit-subunit binding free energy $\gb$ as a function of the simulation well-depth $\eb$ according to $\gb(n)=-\Delta \nci\eb-T(\ssb+\ssc(n))$ where $\Delta \nci$ is the number of subunit-subunit contacts added to form a cluster of size $n$ (assuming the lowest energy configuration at each size $n$), $\ssb(n)$ denotes the translational and rotational binding entropy penalties and $\ssc$ denotes the change in `configurational' entropy  associated with degeneracy of the lowest energy configuration for each cluster size $n$.  The binding entropy can be estimated from a saddle point approximation of the partition function for a subunit dimer as \cite{Hagan2006,Hagan2009}
\begin{align}
\ssb/\kb \approx-\frac{3}{2}\log \left.\frac{\beta \partial^2U_{\text{att}}(r)}{\partial r^2}\right|_{r=2^{1/6}\sigma} -\frac{1}{2}\log\frac{(\beta \eb)^3\pi^7}{\tc^4 \pc^2}
\label{eq:sb}.
\end{align}
We note that this approximation provides an accurate description of the dimerization free energy but neglects additional entropy penalties incurred by subunits forming more than one contact.
The number of contacts $\{\Delta c(n)\}$ and configurational entropy $\{\ssc(n)\}$ values for building an icosahedron are given in Table S2 of Roldao \etal \cite{Roldao2012}. The shifted binding energy (Eq.~\eqref{eq:ebeff}) is modified to
$\ebeff=\eb + \frac{\nn+1}{\nn-1} \log \fa  + 3\log\left(\ebeff/\eb\right)$.

The subunit association rate $k$ is the only adjustable parameter we used in this comparison. While $k$ could be estimated directly from simulations, its value changes for each cluster size due to different occluded volumes neighboring subunits. For simplicity, we use one average value $k=0.003\sigma^3/t_0$, which we estimated (by eye) by comparing master equation and simulation results for $\fc(t)$ for several values of $\eb$.

Although the master equation kinetics do not perfectly match BD results, their agreement is reasonable considering the approximations in our estimate of $\gb$ and the simplifications inherent in the master equation.  For example, in Fig.~\ref{fig:compare} we show the master equation results for $\fc$ calculated at long time $10^7 t_0$ as a function of the binding energy parameter and activation energy. The most significant difference is that the BD simulations result in a small number of completed capsids at large $\eb$ even for \NC, whereas the master equation does not.  We believe this difference arises due to binding of oligomers in the BD simulations \cite{Hagan2006} which is not accounted for in the master equation. We note however that the effect of oligomer binding would diminish at smaller subunit concentrations.

\subsection{Rate equation model with conformation changes}
\label{sec:rateEquationAppendix}

Extension of Eqs.~\eqref{eq:rateEquations} to include interconversion of free subunits between inactive and active conformations results in
\begin{align}
\label{eq:rateEquationsConf1}
\frac{d c_1}{d t} =& -k_1 c_1^2 + 2 \bar{k}_2 c_2 +\sum_{n=2}^{N}-k_n c_n c_1 + \bar{k}_n c_n  \\
 & + \fActive c_1 - \bActive c_1 \nonumber \\
\label{eq:rateEquationsConf2}
\frac{d c_n}{d t} = & \left(k_{n-1} c_1 + k^*_{n-1} \cstar_1\right) c_{n-1}
  - \left(k_n c_1 + k^*_{n} \cstar_1\right) c_n \nonumber \\
 & -\left(\bar{k}_n + \bar{k}^*_n\right) c_n + \left(\bar{k}_{n+1}+\bar{k}^*_{n+1}\right) c_{n+1}   \nonumber \\
 & \mbox{for }n=2\dots N \\
\label{eq:rateEquationsConf3}
\frac{d \cstar_1}{d t} = & \sum_{n=2}^{N}-k^*_n c_n \cstar_1 + \bar{k}^*_n c_n
 - \fActive \cstar_1 + \bActive c_1
\end{align}
where $\cstar_1$ and $c_1$ are respectively the concentrations of inactive and active free subunits, $\fActive$ and $\bActive$ are the active/inactive interconversion rate constants related by $\fActive/\bActive=\exp\left(-\gA/\kt\right)$, and $k^*_n$ and $\bar{k}^*_n$ are the rate constants for association and dissociation of inactive subunits to nucleated partial capsids. For simplicity, we take the nucleus size for inactive subunits binding equal to the energetic critical nucleus size, and the association rate constant for inactive and active subunits to be equal above this size.  Specifically, for \YA, $k^*_n=\Theta(n-\nnuc)f_n$ with $\Theta(n)$ the Heaviside function and $f_n$ the association rate constant for active subunits, while for \NA, $k^*_n=0$ $\forall n$. Alternate choices for these quantities do not qualitatively affect the results. Finally, dissociation rate constants are given by detailed balance, $\bar{k}^*_n=k^*_n \exp\left[(\dgn -\gA)/\kt\right]/v_0$, where the free energy change upon association of an inactive subunit includes the activation energy $\gA$. The initial condition is $\cstar(0)=(1-\fa)c_0$, $c_1(0)=\fa c_0$,  $c_n(0)=0$ for $n>1$.

\subsection{Brownian dynamics model details}
\label{sec:patchyAppendix}
We use the patchy-sphere model presented in Ref.~\cite{Perkett2014}; our description here closely follows that reference. The minimum energy structure is a complete capsid of 20 subunits, which have a spherical excluded volume with three attractive patches, or bond vectors, that are separated by $108^\circ$ and rotate rigidly with the subunit.  The attractive interaction between two complementary bond vectors on respective subunits $i$ and $j$ is maximized when (1) the distance between the attractors $r_{ij}^b$ is minimized, (2) the angle $\theta_{ij}^b$ between bond vectors is minimized, and (3) the dihedral angle $\phi_{ij}^b$ calculated from two secondary bond vectors, which are not involved in the primary interaction, is minimized.  A schematic of the subunit interactions is shown in Fig. \ref{fig:patchy_model}.  Minimizing $\phi_{ij}^b$ creates an interaction that resists torsion and enforces angular specificity commensurate with a complete capsid.  The potentials are given by Eqs \eqref{eq_potential}  \cite{Hagan2008,Perkett2014} \begin{eqnarray}
U &=& U_\text{rep}(R_{ij}) + \sum_b U_\text{att}(r_{ij}^b)S(\theta_{ij}^b, \phi_{ij}^b) \nonumber \\
U_\text{rep}(R_{ij}) &=&  \LJ{12}(R_{ij},2^\frac16 \sigma,\sigma)  \nonumber \\
U_\text{att}(r_{ij}) &=&  \eb \LJ{12}( (r_{ij} + 2^\frac16 \sigma), 2^\frac16 \sigma, \rc ) \nonumber \\
S(\theta,\phi) &=& \frac14 \ \Theta(\theta - \theta_c) \Theta(\phi - \phi_c)  \nonumber \\
        &\phantom{=}&\left( \cos( \pi \theta / \theta_c) + 1\right) \left(\cos(\pi \phi / \phi_c) + 1 \right)
\label{eq_potential}
\end{eqnarray}
with $\LJ{p}$ a generalized truncated and shifted Lennard-Jones function:
\begin{align}
\LJ{p}(x,x_\mathrm{c},\sigma) \equiv &
      4 \left( \left(\frac{x}{\sigma}\right)^{-p} - \left(\frac{x}{\sigma}\right)^{-p/2}  \right. \nonumber \\
        & -\left. \left(\frac{x_\text{c}}{\sigma}\right)^{-p} + \left(\frac{x_\text{c}}{\sigma}\right)^{-p/2} \right) \Theta(x-x_\text{c})
\label{eq_LJ}
\end{align}
In Eq. \eqref{eq_potential} the index $b$ sums over pairs of complementary bond vectors, $\Theta(x)$ is the Heaviside step function and  $R_{ij}$ is the subunit center-to-center distance.


{\bf Conformational dynamics.}
In our BD simulations with conformational dynamics, free subunits stochastically switch between inactive and active conformations. The only difference between the two conformations is their interaction partners.  For \NA, inactive subunits do not experience attractive interactions with any subunit (but they still experience the repulsive excluded volume with all types of subunits), while pairs of active subunits experience the attractive interactions described above. \YA has a similar matrix of interactions, except that in active subunits experience attractive interactions with any subunit in a partial capsid which has at least one completed polygon (meaning a closed cycle of interactions among subunits). In particular, inactive subunits experience no attractions with other inactive subunits, free active subunits, or subunits in partial capsids with no completed polygon, but do experience attractions to partial capsids with at least one completed polygon. Pairs of active subunits experience attractions as described above.
Since we focus on the limit of fast conformational dynamics (relative to assembly timescales), subunits underwent conformational sampling with frequency $2.5t_0^{-1}$. At this frequency, the conformation of each free subunit was stochastically set to inactive or active with respective probabilities $1-\fa$ and $\fa$.

\subsection*{Acknowledgements}
This work was supported by the NIH, Award Number R01GM108021 from the National Institute Of General Medical Sciences and the Brandeis Center for Bioinspired Soft Materials, an NSF MRSEC,  DMR-1420382. Computational resources were provided by the NSF through XSEDE computing resources and the Brandeis HPCC which is partially supported by the Brandeis MRSEC.

\bibliography{all-references_titlecase}

\begin{thebibliography}{89}
\bibitem [{Pin()}]{PintNote}%
  \BibitemOpen
  \href@noop {} {}\bibinfo {note} {Since the master equations can be integrated
  to arbitrarily long times, we report the maximum value of $\pint$ observed at
  any time for a given parameter set.}\BibitemShut {Stop}%
\end{thebibliography}

\end{document}